\documentclass[fleqn,usenatbib]{mnras}

\usepackage{newtxtext,newtxmath}
\usepackage{comment}
\usepackage{array}
\usepackage{float}
\usepackage{stfloats}
\usepackage{tabularray}
\usepackage{tabularx}
\usepackage{threeparttable}
\UseTblrLibrary{booktabs}
\usepackage{lipsum}
\usepackage{multirow}
\usepackage{mathtools}
\usepackage[T1]{fontenc}
\usepackage{ae,aecompl}
\usepackage{graphicx}	
\usepackage{amsmath}	
\usepackage{enumitem} 
\usepackage{xfrac} 
\usepackage{algorithm}
\usepackage{algpseudocode}

\setlist[itemize]{leftmargin=*}

\newcommand{\ud}{\mathrm{d}}

\newcommand{\derfrac}[2]{\frac{\ud #1}{\ud #2}}

\newcommand{\pygad}{\texttt{pygad}}

\newcommand{\frost}{\texttt{FROST}}
\newcommand{\bifrost}{\texttt{BIFROST }}

\newcommand{\mcluster}{\texttt{McLuster}}

\newcommand{\twoa}{\texttt{2MBH\textunderscore e099}}
\newcommand{\twob}{\texttt{2MBH\textunderscore e05}}
\newcommand{\twoc}{\texttt{2MBH\textunderscore e005}}

\newcommand{\onec}{\texttt{1MBH\textunderscore e005}}

\newcommand{\zeroc}{\texttt{noMBH\textunderscore e005}}
\newcommand{\isotwo}{\texttt{2MBH\textunderscore iso}}

\newcommand{\isozero}{\texttt{noMBH\textunderscore iso}}

\title[MBHs in merging star clusters]{The role of Massive Black Holes in merging star clusters: dynamical evolution, stellar \& compact object ejections and gravitational waves}

\author[Souvaitzis et al.]{
Lazaros Souvaitzis$^{1}$\thanks{E-mail: lazaros@mpa-garching.mpg.de}, 
Antti Rantala$^{1}$
and Thorsten Naab$^{1}$ 
\\
$^{1}$Max Planck Institute for Astrophysics, Karl-Schwarzschild-Str. 1, 85748, Garching, Germany\\
}

\date{Accepted XXX. Received YYY; in original form ZZZ}

\pubyear{2025}

\begin{document}
\label{firstpage}
\pagerange{\pageref{firstpage}--\pageref{lastpage}}
\maketitle

\begin{abstract}
Star clusters can interact and merge in galactic discs, halos, or centers. We present direct N-body simulations of binary mergers of star clusters with $M_{\star} = 2.7 \times 10^4 \: \mathrm{M_{\odot}}$ each, using the N-body code \bifrost with subsystem regularisation and post-Newtonian dynamics. We include 500 $\mathrm{M_{\odot}}$ massive black holes (MBHs) in the progenitors to investigate their impact on remnant evolution. The MBHs form hard binaries interacting with stars and stellar black holes (BHs). A few Myr after the cluster merger, this produces sizable populations of runaway stars ($\sim$800 with $v_{\mathrm{ej}} \gtrsim 50 \mathrm{kms^{-1}}$) and stellar BHs ($\sim$30) escaping within 100 Myr. The remnants lose $\sim30\%$ of their BH population and $\sim3\%$ of their stars, with $\sim$30 stars accelerated to high velocities $\gtrsim 300 \mathrm{kms^{-1}}$. Comparison simulations of isolated clusters with central hard MBH binaries and cluster mergers without MBHs show that the process is driven by MBH binaries, while those with a single 1000 $\mathrm{M_{\odot}}$ MBH in isolated or merging clusters produce fewer runaway stars at lower velocities. Low-eccentricity merger orbits yield rotating remnants ($v_{\mathrm{rot}} \sim 3 \mathrm{kms^{-1}}$) , but probing the presence of MBHs via kinematics alone remains challenging. We expect the binary MBHs to merge within a Hubble time, producing observable gravitational-wave (GW) events detectable by future GW detectors such as the Einstein Telescope and LISA. The results suggest that interactions with low-mass MBH binaries formed in merging star clusters are an important additional channel for producing runaway and high-velocity stars, free-floating stellar BHs and compact objects.
\end{abstract}



\section{Introduction}
Numerous escaping stars and compact objects with velocities large enough to escape the galaxy have been detected or proposed in recent years. At the same time the dynamical evolution of star clusters in the presence of massive black holes (MBHs) has been widely studied, while evidence about interacting or merging star clusters is growing. In this section, we first outline the observational evidence of escaping stars and compact objects (Section \ref{sec:1.1}) , then we present evidence for the presence and formation pathways of MBHs in star clusters (Section \ref{sec:1.2}) and the interactions and/or mergers of star clusters (Section \ref{sec:1.3}) .

\subsection{High-velocity escaping stars and compact objects}
\label{sec:1.1}

The leading scenario of stellar ejections from the Galactic Centre (GC) with velocities exceeding the escape velocity of the Galaxy is based on the 'Hills mechanism' \citep{Hills_1988}, i.e., when a stellar binary comes sufficiently close to the supermassive black hole (SMBH) Sgr A* \citep{GRAVITY_2020,EHT_2022} in the GC, it can be tidally disrupted. This refers to the dynamical break up of the binary in analogy to the disruption of a single star by a massive black hole. This results in the capture of one of the stars on a wide eccentric orbit and the ejection of the companion star with extreme velocities of $1000 \: \mathrm{km} \: \mathrm{s^{-1}}$, by far larger than the escape velocity of the Milky Way (MW) . The ejection due to close encounter with the massive BH (MBH) and velocities higher than escape velocity are the two properties that define a hyper-velocity star (HVS). The first HVSs were discovered by \cite{Brown_2005} and \cite{Hirsch_2005} with heliocentric radial velocities of $709 \: \mathrm{km} \: \mathrm{s^{-1}}$ and $708 \: \mathrm{km} \: \mathrm{s^{-1}}$ and at 50 kpc and 19 kpc heliocentric distances respectively, both consistent with GC origin. The very high velocities of those stars allow them to travel long distances following radial trajectories, which makes them ideal probes of the mass distribution of the Galaxy \citep{Gnedin_2005, Kenyon_2008}. In order to reveal their origin, those stars are backwards integrated in a Galactic MW-like potential \citep[e.g.]{Price-Whelan_2017}. The respective trajectory can then be used to probe the shape of the potential and dark matter halo profile \citep{Gnedin_2005, Yu_2007, Perets_2009,Contigiani_2019}. An additional confirmation of the Hills mechanism was the discovery of a HVS with an extreme velocity of $1800 \: \mathrm{km} \: \mathrm{s^{-1}}$ and a backwards trajectory pointing towards the GC and Sgr A* \citep{Koposov_2019}.  The discovery of the first HVS has lead to numerous dedicated searches \citep{Li_2011, Zheng_2014, Huang_2017, Du_2019, Kreuzer_2020}, reporting a large number of HVSs candidates in the Galactic halo (review in \cite{Brown_2015} and references therein) but the true origin for many of those still remains a mystery. An additional channel for the production of HVSs was proposed by \cite{Begelman_1980Natur}, and \cite{Yu_2003}, extending the classical Hills mechanism: the interaction of single or binary stars with a binary SMBH.  Alternatively, the inspiral of an intermediate mass (in the range of $100 \: \mathrm{M_{\odot}} \lesssim M_{\bullet} < 10^6 \: \mathrm{M_{\odot}}$) black hole (IMBH) towards the GC \citep{Baumgardt_2006, Sesana_2007, Sesana_2009,Wang_YH_2018,Rasskazov_2019,Evans_2023} could also explain the origin and ejection mechanism of HVSs.

Another class of high-velocity stars is observed to have large peculiar velocities, in the range of $40 \: \mathrm{km} \: \mathrm{s^{-1}} \leq v_{\mathrm{pec}} \leq 200 \: \mathrm{km} \: \mathrm{s^{-1}}$ \citep{1961BAN....15..265B,Gies_1986, Hoogerwerf_2001,Perets_2012}, commonly called 'runaway stars' (RASs). Those are usually early type (O-type, B-type) stars with a Galactic disk origin. Their ejection mechanisms are usually divided in two types \citep{Blaauw_1993, Hattori_2019, Carretero_Castrillo_2023} : (a) the dynamical ejection mechanism (DEM), where strong 3- and 4- body encounters between stars and/or BHs in dense stellar environments lead to the ejection of one of the members \citep{Poveda_1967, Aarseth_1972, Hut_1983, Ryu_2017a,Ryu_2017b, Weatherford_2023} and (b) the binary ejection mechanism (BEM), where primary of the binary members undergoes a supernova explosion (SN) and the secondary (the runaway star) moves on with close to its original orbital velocity. The primary remnant receives a kick from the SN which might unbind it from the secondary \citep{Zwicky_1957, Blaauw_1961, Portegies_Zwart_2000, Justham_2008, Pakmor_2013, Neunteufel_2020, Rajamuthukumar_2024} . \cite{Heber_2008} have discovered an unbound star travelling with heliocentric radial velocity of $>400 \: \mathrm{km} \: \mathrm{s^{-1}}$ with a Galactic disk origin and have classified it as a hyper-runwaway star (HRS) . Finally, alternative mechanisms for the origin of high-velocity stars include the encounters with satellite galaxies and nearby galaxies such as the Large Magellanic Cloud (LMC) \citep{Gualandris_2007,Boubert_2017,Erkal_2018} and the Andromeda Galaxy \citep{Sherwin_2008} and from star clusters tidally interacting with a single or a binary SMBH \citep{Capuzzo_Dolcetta_2015,Fragione_2016,Fragione_2017}.

Our knowledge on stellar populations in the Milky Way has been vastly expanded by the $Gaia$ satellite mission of the European Space Agency (ESA) and the unprecedented quality of astrometric and photometric data it provides. Already from the first data release (DR1) \citep{Gaia_2016} , the number of HVS candidates has significantly increased \citep{Marchetti_2018} , contributing at least 14 more objects with a total velocity in the Galactic rest frame of $>400 \: \mathrm{km} \: \mathrm{s^{-1}}$ and one classified as a HVS. A crucial step towards the improvement of modern astrometry came with the second data release (DR2) \citep{Gaia_2018} providing us positions and proper motions of more than one billion stars and the radial velocities for about 7 million of those. The number of HVS candidates reported in the literature before Gaia's DR2 was close to 500 \citep{Hirsch_2005, Brown_2006, Brown_2008, Brown_2012, Brown_2014, Li_2011,Li_2015, Huang_2017, Vennes_2017}, with about 20 of them being faint and blue stars located in the Galactic halo with very high radial velocities and being classified as HVSs. Utilizing Gaia DR2 \cite{Hattori_2018} found 30 $(>480 \: \mathrm{km} \: \mathrm{s^{-1}})$ old metal-poor stars, with 2-3 having LMC and 1-2 GC origin. Three white dwarfs (WDs) with extreme velocities of $1000  \: \mathrm{s^{-1}} \leq v \leq 3000 \: \mathrm{km} \: \mathrm{s^{-1}}$, were also discovered in DR2 from \cite{Shen_2018}. \cite{Marchetti_2019} found 20 stars unbound ($> 80 \% $ probability) from the Galaxy, seven of which are consistent with a GC origin and 13 with an origin outside of the Milky Way. By combining Gaia DR2 and the 7th data release of the Large Sky Multi-object Fiber Spectroscopic Telescope (LAMOST), \cite{Li_2021} reported a total of 591 high-velocity candidates with 43 of them having more than $50 \%$ probability of being unbound from the Galaxy, increasing the number of known candidates by a factor of two. Despite the additional proper motions and radial velocities of 34 million stars that the third Gaia data release (DR3) \citep{Gaia_2023} has provided, \cite{Marchetti_2022} found no additional  HVS candidates $>400 \: \mathrm{km} \: \mathrm{s^{-1}}$ and \cite{Liao_2023} reported only two with radial velocities larger than $500 \: \mathrm{km} \: \mathrm{s^{-1}}$ without strong evidence for a GC origin. Finally, the recent discovery \citep{Huang_2024} of high-velocity star (J0731+3717) with $v_{\mathrm{ej}} \approx 548 \: \mathrm{km} \: \mathrm{s^{-1}}$, whose backward trajectory $21 \: \mathrm{Myr}$ ago, reveals its origin from MW globular cluster M15.

\subsection{Evidence for massive black holes in star clusters}
\label{sec:1.2}

The existence and formation pathway of IMBHs remains an open question. Dense stellar environments like dwarf galaxies and massive star and globular clusters are ideal sites for their formation and growth \citep{Askar_2024}. Although limited, observational evidence of IMBHs have been growing recently \citep[see e.g.][and references therein]{Mezcua_2017, Greene_2020}{}{}. One of the most recent and clear piece of evidence to date, is the detection GW190521 \citep{Abbott_2021} of an IMBH with $M_{\bullet} \sim 150 \: \mathrm{M_{\odot}}$ via gravitational waves (GW) from two coalescing stellar-mass black holes of $m_{\bullet} \sim 66 \: \mathrm{M_{\odot}}$ and $m_{\bullet} \sim 85 \: \mathrm{M_{\odot}}$ each.  Apart from GWs numerous candidates of accreting IMBHs have been proposed due to X-ray emission detected in galaxies nearby, in the mass range of  $200 \lesssim M_{\bullet} \lesssim 10^5 \: \mathrm{M_{\odot}}$ \citep{Matsumoto_2001,Strohmayer_2003,Farrell_2009,Mezcua_2013,Mezcua_2015,Mezcua_2017}. Moreover, an IMBH of $M_{\bullet} \sim 10^4 \: \mathrm{M_{\odot}}$ was recently proposed \citep{Wen_2021} as a the X-ray source of a tidal disruption event (TDE) in a massive object of $M_\mathrm{\star} \sim 10^7 \mathrm{M_{\odot}}$ \citep{Lin_2018}, which corresponds either to a stripped galactic nucleus or a globular cluster. 

Evidence for the existence of IMBHs or central populations of stellar BHs in various stellar and globular clusters is supported by dynamical signatures from observations and dynamical modeling. However, making  a clear distinction between the two alternatives remains challenging. Among the top candidates of galactic (MW) globular clusters hosting an IMBH are $\omega$Centauri ($\omega$Cen) and 47 Tucanae (47 Tuc). For the former dynamical models \citep{van-der-Marel_2010,Noyola_2010,Jalali_2012,Baumgardt_2016} have proposed a central IMBH in the $1.2 \times 10^4 \: \mathrm{M_{\odot}} \lesssim M_{\bullet} \lesssim 5 \times 10^4 \: \mathrm{M_{\odot}}$. Only recently observations of proper-motions of fast-moving stars in the inner $0.08 \: \mathrm{pc}$ of $\omega$Cen from HST have been analyzed by \cite{Haeberle_2024}, who found that the excess velocities can be explained by the presence of an IMBH with a corresponding lower mass limit of $M_{\bullet} \sim 8.8 \times 10^3 \: \mathrm{M_{\odot}}$. On the other hand, the situation in 47 Tuc is still under debate. X-ray \citep{Grindlay_2001} and radio \citep{de-Rijcke_2006} data observations have provided upper mass limits of $\sim 470 \: \mathrm{M_{\odot}}$ and $\sim 2060 \: \mathrm{M_{\odot}}$ for a central IMBH in 47 Tuc, while recent dynamical modeling \citep{Kiziltan_2017a,Kiziltan_2017b} supports this evidence predicting a mass of $M_{\bullet} \sim 2300 \: \mathrm{M_{\odot}}^{+1500}_{-800}$. Further observations of millisecond pulsars (MPS) have shown that the presence of an IMBH is not required to explain the data \citep{Freire_2017,Abbate_2018} , in line with multi-mass dynamical modeling results \citep{Mann_2019,Henault-Brunet_2020}. Updated dynamical models by \cite{DellaCroce_2023} place an upper limit of $\sim 578 \: \mathrm{M_{\odot}}$ and only the latest ultradeep Australia Telescope Compact Array (ATCA) have revealed a central compact radio source associated with a faint X-ray emission corresponding to an IMBH of $M_{\bullet} \sim 54-6000 \: \mathrm{M_{\odot}}$. 

Numerous studies have been conducted providing additional evidence of a central IMBH or a dark central cluster \citep{Ibata_2009,Kamann_2014,Baumgardt_2016,Nguyen_2017,Gieles_2018,Vitral_2022} in such environments. A detailed analysis of 3D spectroscopic data by \cite{Kamann_2016} revealed the presence of a $M_{\bullet} \sim 600  \: \mathrm{M_{\odot}}$ IMBH in the core-collapsed cluster NGC6397, while \cite{Rui_2021} suggested a central sub-system of stellar remnants supported by \cite{Vitral_2022} who utilized Gaia and HST proper motion data, consistent with a dark concentration mass of $ \sim 1000 \: \mathrm{M_{\odot}}$. A less debated case of whether the central mass corresponds to a single IMBH or to a sub-system of dark objects is that of NGC6388. Combining spectroscopic data from the \textit{Very Large Telescope} (VLT) and HST, \cite{Luetzgendorf_2011} proposed an IMBH of $M_{\bullet} \sim (1.7 \pm  0.9) \times 10^4  \: \mathrm{M_{\odot}}$ confirmed by \cite{Lanzoni_2013} who placed an upper limit of  $ \sim 2000 \: \mathrm{M_{\odot}}$ for the mass of the IMBH. Finally, similar findings have been proposed for Andromeda (M31) galaxy. For example, with the use of \textit{Hubble Space Telescope} (HST) data, \cite{Gebhardt_2002} reported the detection of a $M_{\bullet} \sim 2 \times 10^4  \: \mathrm{M_{\odot}}$ IMBH in G1 cluster of M31 galaxy. Dynamical models by \cite{Baumgardt_2003} proposed that the presence of an IMBH in G1 is not needed to explain the observed data, while X-ray emission detected by \textit{XMM-Newton} could not distinguish between an accreting IMBH or a low-mass X-ray binary \citep{Pooley_2006}. Additionally, there has been recent evidence for a $M_{\bullet} \sim 10^5  \: \mathrm{M_{\odot}}$ IxsMBH in G078 \citep{Pechetti_2022}, which is the most massive globular cluster of M31, originating from the tidally stripped nucleus of a dwarf galaxy \citep{Fuentes-Carrera_2008}.


One of the leading mechanisms for the formation of an IMBH, is the growth of low-mass BHs ($m_{\bullet} < 50 \: \mathrm{M_{\odot}}$) in dense stellar environments via successive stellar and compact object collisions \citep{Stone_2017,Arca_Sedda_2021,Rizzuto_2020,Rizzuto_2022,Arca_Sedda_2023_dragonsim_II}. The high densities in the centre of such environments leads to a sequence of mergers capable of producing BHs on the intermediate mass range $100 \: \mathrm{M_{\odot}} \lesssim M_{\bullet} < 10^4 \: \mathrm{M_{\odot}}$ \citep{Atallah_2023,Arca_Sedda_2021,Fragione_2022,Mapelli_2021,Rizzuto_2022}. Such merger products can receive velocity kicks ranging from tens up to $5000 \: \mathrm{km} \: \mathrm{s^{-1}}$ \citep{Campanelli_2007,Lousto_2019} due to the GW-induced recoil. Consequently, it is rather unlikely that the final product can be retained in low escape velocity environments such as stellar and globular clusters \citep{Gerosa_2019}. The extreme densities of nuclear star clusters (NSCs) combined with the increased escape velocities $v_{\mathrm{esc}} \gtrsim 100 \: \mathrm{km} \: \mathrm{s^{-1}}$ \citep{Neumayer_2020}, makes them ideal sites for the retention of IMBHs. The hierarchical assembly of massive star clusters and NSCs makes retaining the merger products more likely as the black holes end up residing in environments with higher escape velocities than their original birth clusters \citep{Rantala2024}.

The theory of pair-instability (PPSN) and pulsation pair-instability supernovae (PSN) predicts a mass gap on BH formation ranging between $\sim \: 50-130 \: \mathrm{M_{\odot}}$ , due to single stellar evolution only \citep{Fowler_1964,Woosley_2007,Woosley_2017}. In young and massive star clusters, a star can grow to very high masses (very massive star with $m_{\star} > 150 \: \mathrm{M_{\odot}}$, therefore VMS) via repeated stellar collisions which collapses to an IMBH within or above the mass gap, which may further grow through accretion during TDEs. This so-called \textit{fast runaway collision} scenario, has been numerically studied \citep{Portegies_Zwart_1999,Portegies_Zwart_2002,Freitag_2006a,Freitag_2006b,Rizzuto_2020,Rizzuto_2023,Arca_Sedda_2023_dragonsim_I,Arca_Sedda_2023_dragonsim_II,Arca_Sedda_2023_dragonsim_III,Prieto_2024,Rantala2024} where VMSs are efficiently formed, resulting in MBHs of $M_{\bullet} \gtrsim 10^2 - 10^4 \: \mathrm{M_{\odot}}$ when they collapse. Finally, it is important to mention, that the mass-loss from VMSs is still unclear and further research in stellar evolution models is needed to understand if and how long such stars could survive even if they do form \citep{Sabhahit_2022,Sabhahit_2023}.

The formation of IMBHs and the co-evolution of massive star cluster hosts has been studied mostly in isolated numerical setups \citep{Wang_2016,Rizzuto_2023,Arca_Sedda_2023_dragonsim_I} and only recently in full hierarchical by \cite{Rantala2024}. Recent observations by James Webb Telescope (JWST) suggest the formation of clumped clusters like the Cosmic Grapes at $z>6$ redshift \citep{Fujimoto_2024}. Furthermore, five 
massive ($M_{cl} \sim 10^6 \; \mathrm{M_{\odot}}$)  star clusters, the so-called Cosmic Gems, have been detected by JWST at $z=10.2$, providing additional evidence to such hierarchical assembly. This is also supported by previous observations of star cluster forming regions \citep[e.g.][]{Zhang_2001,Bastian_2005,Grasha_2017,Menon_2021}{}{} and simulations of star-burst environments \citep{Lahen_2020}, where a monolithic collapse seems unlikely.

The conditions required to produce massive clusters, such as globular clusters, occur in interacting galaxies and star forming regions during the collapse of giant molecular clouds \citep[see e.g.][and references therein]{Krumholz_2019}{}{}. The combined effects of the properties of the collapsing cloud and the tidal field from the forming galaxy may lead to efficient angular momentum transport in the star cluster forming region, resulting in rotating clusters \cite{Lahen_2020}. Although debated in the past, there is clear observational evidence nowadays of rotating GCs in the Milky Way \citep{Bellazzini_2012, Fabricius_2014} with rotational velocities of the order of a few $\mathrm{km \: s^{-1}}$. Rotation can have significant effects on the overall evolution \cite{Fiestas_2008} of a star cluster, for example spinning up its core and increasing the stellar ejection rate \citep{Einsel_1999, Ernst_2007} or when combined with stellar evolution of the system \cite{Kamlah_2022}. 

Similar to the coupled formation and evolution of IMBHs and massive star clusters, the properties of NSCs and SMBHs are closely related to those of their host galaxies \citep{Ferrarese_2006,Leigh_2015,Capuzzo_Dolcetta_2017} which indicates that the formation and evolution of the two is tightly related \citep{Antonini_2015,Neumayer_2020}. Two scenarios have been proposed for the formation of NSCs. Galactic nuclei with sufficiently high gas densities can trigger in-situ star formation \citep{Loose_1982,Mihos_1994,Milosavljevic_2004,Nayakshin_2007,Aharon_2015} leading to the formation of a NSC, or by cluster inspirals and mergers towards the galactic centre due to dynamical friction \citep{Tremaine_1975,Capuzzo_Dolcetta_1993,Loose_1982,Agarwal_2011,Tsatsi_2016}. The latter suggests that clusters may interact with each other during their infall towards the GC. In conclusion though, the diversity of stellar age and metallicity observed in galactic nuclei implies that both scenarios contribute to the overall evolution of NSCs \citep{Antonini_2015,Guillard_2016,Neumayer_2020,Do_2020,Arca_Sedda_2020} . 

\subsection{Binary star clusters: interactions and mergers}
\label{sec:1.3}

The interaction and merging of star clusters is theoretically expected and observed during their formation \citep[see e.g.][]{deLaFuenteMarcos_2009,Lahen_2020}. Especially for LMC, it has been found that the fraction of binary star clusters is roughly $\sim 12\%$ \citep{Dieball_2002} , while a similar number $(\sim 10\%)$ is expected for the Milky Way \citep{deLaFuenteMarcos_2010}. \cite{Bhatia_1991} published a catalogue with numerous binary star cluster candidates in LMC, but their formation path for most of them remains unclear. Apart from forming together at birth, star clusters undergo tidal interactions and mass exchange at later evolutionary phases in the disks of galaxies \citep{Khoperskov_2018,Mastrobuono_Battisti_2019,Camargo_2021,Ishchenko_2024a}. Recent high resolution observations  of LMC, provided evidence of pairs of star clusters in collision course due to tidal capture \citep{Mora_2019,Giusti_2023}. The kinematic properties of almost the entire population of the MW globular clusters is nowadays available thanks to the Gaia DR3 release \citep{Gaia_2018,Gaia_2023}. A detailed investigation on 16 Galactic binary cluster candidates from Gaia DR3 \citep{Gaia_2023} was recently done by \cite{Angelo_2022} who found 4 pairs of bound open clusters. \cite{Ishchenko_2023a} conducted a statistical analysis for the interaction probability of the galactic globular clusters utilizing the Gaia DR2 release \citep{Gaia_2018} and found a significant rate of close encounters and an average of 10 intersecting with-each-other trajectories per cluster. Additionally, \cite{Chemerynska_2022} integrated the orbits of the 150 clusters in a static MW-like potential up to $5 \: \mathrm{Gyr}$ and found a probability above $20\%$ for cluster collisions.  Finally, recent studies have shown that many globular clusters inhibit disk-like kinematics  \citep{Casetti_Dinescu_2010,VandenBerg_2013} providing evidence of close orbital passages and angular momentum gain leading to rotation. This leads to the formation of IMBH binaries potentially leading to GW-driven coalescence.

\vspace{0.5cm}
In this paper we investigate the presence of single and binary MBHs in star clusters as the potential origin for high velocity stars. We use direct N-body simulations of idealised isolated and merging star clusters with and without central massive black holes (MBH).  We find that, in particular in the presence of MBH binaries, stars and compact objects (COs) can be accelerated to velocities up to 800 $\mathrm{km} \: \mathrm{s}^{-1}$. The paper is structured as follows: in Section \ref{section: ejections} we describe the various mechanisms leading to the production of high velocity stars and COs.  The simulation methods and initial conditions are presented in Section \ref{sec:methods}, while the results about cluster evolution, the formation and evolution of MBH binaries and the demographics of ejections in the Sections \ref{sec:clusters}, \ref{sec:MBHB_evol}, \ref{sec:ejection} respectively. Finally, we summarize our conclusions in Section \ref{sec:Conclusions} .

\section{Dynamical Ejection Mechanisms (DEM) in Star $\&$ Globular Clusters}\label{section: ejections}
In this section we review the basic DEM processes operating in star clusters, where a single or binary MBH and/or a sub-cluster of stellar-mass BHs are present \citep{Kulkarni_1993, Quinlan_1996, O_Leary_2006, Sesana_2006, Trenti_2007,Morscher_2015, Fragione_2016,Fragione_2019, Subr_2019, Rasskazov_2019, Weatherford_2023}. If runaway and hyper-velocity stars are observed and a star cluster origin is confirmed, it would be smoking gun evidence for the existence of massive black hole seeds in the heart of such systems. Additionally, understanding their dynamical production channels can be used to disentangle the tension between the presence of a single massive or a collection of stellar-mass 'dark' objects (BHs) in the cores of massive clusters \citep{Anderson_2010,Noyola_2010,Lu_2011,Strader_2012,Chomiuk_2013,Feldmeier_2013,Peuten_2016,Zocchi_2017,Zocchi_2018,Bellini_2018,Vitral_2022,Haeberle_2024,Paduano_2024,Huang_2024} .

\subsection{Single Stellar Encounters and 2-Body Relaxation}
The gravitational interaction between two stars in the vicinity of a BH, with masses $m_1$ and $m_2$, results in a change in the velocity of the $m_1$ \citep{Yu_2003} ,

\begin{equation}
\begin{aligned}
\delta v_1 & =\frac{2 G m_2}{\left[G^2\left(m_1+m_2\right)^2 / v_{1 2}^2+b^2 v_{1 2}^2\right]^{1 / 2}} \\
& \leq \frac{2 G m_2}{\left[2 G\left(m_1+m_2\right) b\right]^{1 / 2}} \\
& =440 \mathrm{~km} \mathrm{~s}^{-1}\left(\frac{2 m_2}{m_1+m_2}\right)^{1 / 2}\left(\frac{m_2}{1 \mathrm{M}_{\odot}}\right)^{1 / 2}\left(\frac{1 \mathrm{R}_{\odot}}{b}\right)^{1 / 2}
\end{aligned}
\end{equation}
where $v_{1 2}$ is the relative velocity of the two stars and $b$ is the impact parameter \citep{Binney2008}. Although rare, an encounter of this type with small enough impact parameter can result in significant velocity kick. For example, if $m_1=m_2=1 \mathrm{M_{\odot}}$ and $b=10^{-6} \mathrm{pc}$, then $\delta v_1 \geq 30 \mathrm{~km} \mathrm{~s}^{-1}$. 

The collective motion of objects in a N-body system, induces time-varying perturbations on the background potential of the system. A non-smooth potential introduces energy and angular momentum exchange between the cluster members, causing them to randomly diffuse in the phase space. The net effect of all the uncorrelated 2-body encounters can lead to a change of each body's velocity by order of itself $\Delta v / v \sim 1$ \citep{Chandrasekhar_1942, Spitzer_1987, Heggie_2003, Aarseth_2008}, allowing an object to escape within the relaxation timescale and velocity of the order $v_{2b} \approx 2\sqrt{\sigma}$, where $\sigma$ is the velocity dispersion the cluster. A star located in the outskirts of a cluster with local velocity dispersion $\sigma \approx 2 \: \mathrm{km} \: \mathrm{s}^{-1}$, could then escape through that process with $v_{2b} \approx 2.8 \: \mathrm{km} \: \mathrm{s}^{-1}$.

\subsection{Encounters of Singles with Binary Stars}
Interactions between single and binary stars are common in star clusters and play a key role to binary evolution \citep{Heggie_1975}. Those interactions can lead to the formation of new binaries from single ones or through the exchange of binary members \citep{Valtonen_2006}. Lower mass objects are accelerated to high velocities when encounters between unequal mass stars take place \citep{Spitzer_1987}, potentially leading to their escape from the host cluster. The critical velocity of a star (member of the cluster) capable of ionizing (the process where an encounter leads the break up of a bound system to its individual components) the binary in the centre-of-mass of a triple system is given by \citep{Hut_1983, Hills_1975b, Sigurdsson_1993}  
\begin{equation}
\begin{aligned}
v_c & =\sqrt{G \frac{m_1 m_2}{m_3} \frac{\left(m_1+m_2+m_3\right)}{\left(m_1+m_2\right)} \frac{1}{a}} \\
& =\sqrt{\frac{G m_{123}}{a} \frac{\mu_{12}}{m_3}},
\end{aligned}
\end{equation}
Here $\mu_{12} = m_1 m_2 / (m_1 + m_2)$ , $m_{123} = m_1 + m_2 + m_3 $ and $a$ is the semi-major axis of the binary with $m_1 \geq m_2$, where the least massive body being the most likely to escape. For a binary with $m_1= 2 \: M_{\odot}$, $m_2= 1 \: M_{\odot}$ and $a= 2 \: \mathrm{AU}$, a third star of $m_3 = 1 \: M_{\odot}$ would need $v_c = 33.8 \: \mathrm{km} \: \mathrm{s}^{-1} $ to ionize the binary. The stochastic nature of such process makes it very challenging to  predict the ejection velocity of the escaping body from generic initial conditions, but numerous detailed numerical 3-body scattering experiments have been conducted \citep[e.g.][]{Hut_1983, Hills_1975b, Sigurdsson_1993}{}{} for the equal-mass case and recently from \cite{Forastier_2024} for unequal masses. Finally, \cite{Samsing_2017} have investigated the role of stellar tides and post-Newtonian (PN) dynamics on such encounters and found that including those effects lead to higher stellar coalescence rates.

\subsection{Binary Encounters with a Single Massive Black Hole} 
The strong tidal forces a star experiences once exerted by an SMBH, are capable of tearing it apart once it gets sufficiently close. This tidal disruption event (TDE) takes place when the orbital pericentre distance $r_{\mathrm{peri}} \leq r_t$, where $r_t$ is the tidal radius given by \citep{Hills_1975}

\begin{equation}
    r_t \approx\left(\frac{M_{\mathrm{SMBH}}}{m}\right)^{1 / 3} r_{*}.
\end{equation}

Consider now a binary system with mass $m_{\mathrm{b}}$ and semi-major axis $a_{\mathrm{b}}$ approaching an SMBH. The same process (replacing $r_{*}$ with $a_{\mathrm{b}}$ and $m=m_{\mathrm{b}}$) will lead to the tidal break-up of the binary, where one member is captured by the SMBH and the other ejected with $v_{\mathrm{ej}}$ \citep{Hills_1988}. This happens when it passes closer to the SMBH than the tidal radius, i.e., when 

\begin{equation}
a_b\left(\frac{M_{\mathrm{SMBH}}}{m_1+m_2}\right)^{1 / 3} \simeq 10 \mathrm{AU}\left(\frac{a_b}{0.1 \mathrm{AU}}\right)\left[\frac{M_{\mathrm{SMBH}}}{10^6\left(m_1+m_2\right)}\right]^{1 / 3} .
\end{equation}

For a star cluster with an IMBH at its centre and a number of binaries inside its core, the above mechanism may efficiently produce high-velocity stars escaping the cluster. During the tidal breakup, the stars receive a velocity change of the order of the velocity relative to the centre of mass of the binary, which for $m_1$ in this case is \citep{Hills_1988, Yu_2003}, 

\begin{equation}
\begin{aligned}
\delta v_1 & \approx \sqrt{\frac{G\left(m_1+m_2\right)}{a_{\mathrm{b}}}}\left(\frac{m_2}{m_1+m_2}\right) \\
& \approx 67 \mathrm{~km} \mathrm{~s}^{-1}\left(\frac{2 m_2}{m_1+m_2}\right)^{1 / 2}\left(\frac{m_2}{1 \mathrm{M}_{\odot}}\right)^{1 / 2}\left(\frac{0.1 \mathrm{AU}}{a_{\mathrm{b}}}\right)^{1 / 2} , 
\end{aligned}
\end{equation}
or in a more practical form by \cite{Bromley_2006}, the velocity of the ejected star is,

\begin{equation}
\label{eq:Hills}
v_{\mathrm{ej}} \approx 460\left(\frac{a_{\mathrm{b}}}{0.1 \mathrm{AU}}\right)^{-1 / 2}\left(\frac{m}{2 \mathrm{M}_{\odot}}\right)^{1 / 3}\left(\frac{M_{\mathrm{IMBH}}}{10^3 \mathrm{M}_{\odot}}\right)^{1 / 6} \mathrm{~km} \mathrm{~s}^{-1} .
\end{equation}

Using Eq. \ref{eq:Hills}, for a binary of $m_{\mathrm{b}}=m=1\mathrm{M_{\odot}}$ approaching an IMBH of $M_{\mathrm{IMBH}}=1000 \: \mathrm{M_{\odot}}$, the ejection velocity can vary from $v_{\mathrm{ej}} \approx 136 \mathrm{~km} \mathrm{~s}^{-1}$ for $a_{\mathrm{b}} = 0.1 \mathrm{mpc}$ up to $v_{\mathrm{ej}} \approx 1360 \mathrm{~km} \mathrm{~s}^{-1}$ for $a_{\mathrm{b}} = 0.01 \mathrm{mpc}$ . For star clusters with large binary fractions, this channel acts as the major contribution to the production of RAs and HVSs \citep{Subr_2019, Fragione_2019, Weatherford_2023}. Finally, \cite{Ryu_2023a} showed by means of hydrodynamics simulations that in the special case where a tight stellar binary interacts with a stellar-mass BH, both stars can be tidally disrupted if the impact parameter is small enough or end up with one of the members becoming unbound (\textit{micro-Hills} mechanism). If the binary consists of a star and a BH, then the stellar ejection could also be accompanied by the formation of a binary BH via binary member exchange \cite{Ryu_2023b}.

\subsection{Encounters with (Massive) Black Holes Binaries} 
\label{sec:MBHB_encounters}
Star clusters are expected to host a population of binary BHs and compact objects \citep{Kulkarni_1993,Portegies_Zwart_2000a,Downing_2010,Tanikawa_2013,Morscher_2015,Rodriguez_2015,Park_2017,Anagnostou_2020,Torniamenti_2022}. The presence of bodies with a range of masses in a star cluster lead to the exchange of kinetic energy and ultimately more massive objects tend to shrink to the centre of the cluster. During this stage, the density of the star cluster core is increased leading to the efficient dynamical formation of stellar and stellar remnant binaries. Encounters between single stars and COs with BH binaries can lead to sufficiently high velocity kicks and result in ejections. Especially when their semi-major axis $a_{\mathrm{BHB}}$ becomes small enough, such that $a_{\mathrm{b}} \leq a_{\mathrm{h}} \equiv G M_2 / 4 \sigma_{c}^2$ \citep{Sesana_2006}, where $\sigma_c$ is the local dispersion velocity \citep{Quinlan_1996}. 
Additionally, a single star approaching a black hole binary (BHB) can be tidally disrupted and further harden the binary up to $20 \%$ for retrograde encounters \citep{Ryu_2022}. The velocity kick of a single or multiple encounters of a star or CO with a BHB is then given by \citep{Yu_2003}, 

\begin{equation}
\label{eq:BHB}
\delta v \approx 1500 \mathrm{~km} \mathrm{~s}^{-1}\left(\frac{2 M_2}{M_1+M_2}\right)^{1 / 2}\left(\frac{M_2}{10^6 \mathrm{M}_{\odot}}\right)^{1 / 2}\left(\frac{1 \mathrm{mpc}}{a_{\mathrm{b}}}\right)^{1 / 2}.
\end{equation}

During the hierarchical formation of a massive star cluster \citep{Gieles_2011,Rantala2024}, lower mass seed BHs can also form binaries. Those BH binaries are usually more massive than stellar remnants, since they can grow by successive mergers or as by products of runaway collisions \citep{Arca_Sedda_2023_dragonsim_II,Arca_Sedda_2023_dragonsim_III,Fujii_2024}. Moreover, similar to how galaxies merge \citep[e.g.][]{Milosavljevic_2001}{}{} leading to the formation of SMBH binaries, galactic globular clusters can also interact and merge. If such clusters contain a massive MBH, their merger remnant becomes the host of an MBH binary, which can then scatter single and binary stars around resulting to their ejection from the cluster. Finally, an MBH-host star or globular cluster on an inspiral orbit towards the galactic centre captured by the SMBH \citep[e.g.][]{Baumgardt_2006,Sesana_2007,Sesana_2009}{}{}, may lead to the formation of an SMBH-IMBH binary which can eject stars at high speed \citep{Rasskazov_2019} once it gets hard ($a_{\mathrm{b}}<a_{\mathrm{h}}$). 

In the special but common in dense stellar environments like globular clusters and the galactic centre case where all interacting bodies are BHs, single-binary encounters in star clusters can efficiently lead to the production of GW-driven merger events \citep{Samsing_2018a, Samsing_2018b, Samsing_2018c}, where one of the BHs receives a strong velocity kick enough to be ejected from the cluster.

\begin{figure*}
  \centering
   \includegraphics[width=\textwidth]{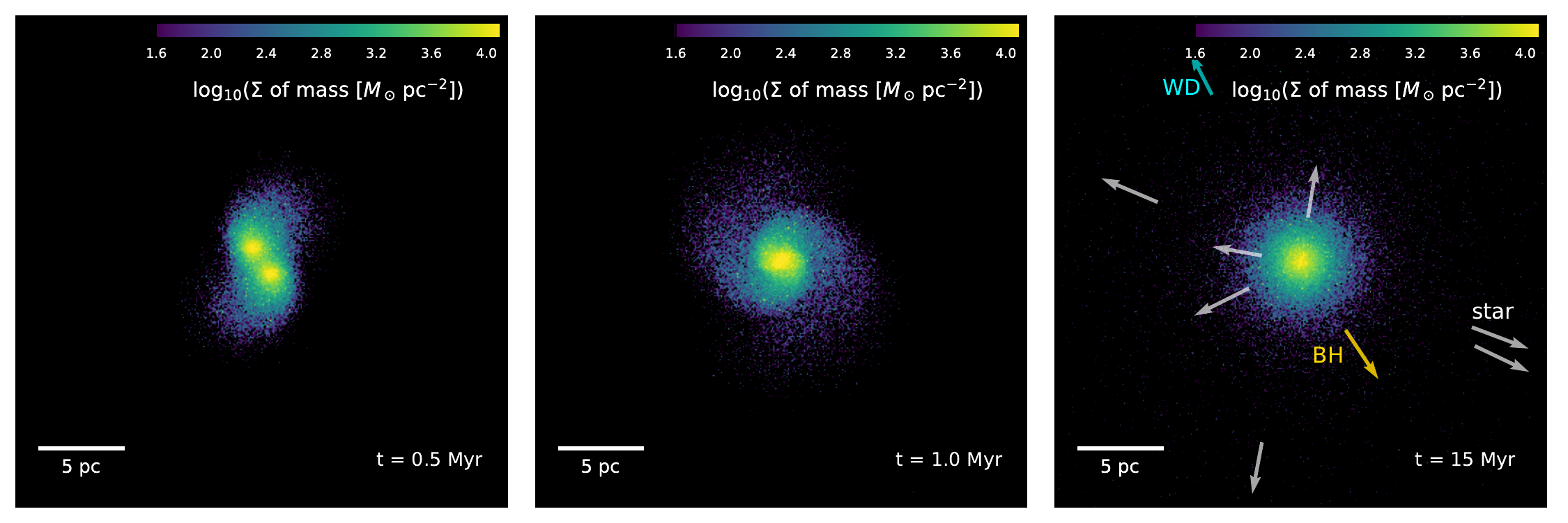}
    \caption{Time sequence of the stellar surface density distribution of a typical star cluster merger simulations with an orbital eccentricity $\mathrm{e}=0.05$ and an MBH in each cluster centre. We show the simulation at an early interacting phase (left panel, t = 0.5 Myr), at the time of the merger of the cluster cores (middle panel, t = 1 Myr) and after the central MBHs have formed a hard binary in the centre of the merger remnant (right panel, t = 15 Myr). By interactions with the MBH binary stars (star, white arrows), stellar BHs (BH, orange arrow) and white dwarfs (WD, cyan arrow) can be ejected at high velocities after interacting with the MBH binary (a few examples are shown). The arrow length scales with velocity and the BH has a velocity of $v_{\mathrm{ej}} \approx 80 \mathrm{~km} \mathrm{~s}^{-1}$.}
    \label{fig:initial_final_cl}
\end{figure*}

\section{Methods}
\label{sec:methods}
To investigate the influence of (MBHs) on the merging of stellar clusters and the resulting population of escaping stars and compact objects (i.e. white dwarfs, neutron stars and black holes), we conduct a series of direct N-Body simulations. These simulations involve merging pairs of stellar clusters, where either both, only the primary or none of them hosts an MBH at their centre. The systems are evolved for 100 million years (Myr) using the direct N-Body code \bifrost{} \citep{Rantala2023},  which we describe briefly in Subsection \ref{section:n-body}. 


\subsection{N-Body Code}
\label{section:n-body}
Throughout this study we used the GPU-accelerated direct-summation N-Body code \bifrost{} \citep{Rantala2023}, which is the updated version of the \frost{} code \citep{Rantala2021}. The code employs a fourth-order forward symplectic integrator (FSI) technique \citep{CHIN1997344, Chin2005, Dehnen2017a}, using a hierarchical implementation \citep{Rantala2021}. Beyond traditional Newtonian accelerations, FSI incorporates gradient accelerations to cancel second-order error terms, resulting in a fourth-order time integration cycle with strictly positive sub-steps. This method was found to be more effective \citep{Chin2007} e.g. for the Kepler problem than traditional Yoshida-type symplectic integrators \citep{Yoshida1990} which use negative sub-steps for higher than second integration order. Furthermore, within the hierarchical integration approach, rapidly evolving components of the simulated systems decouple from slowly evolving components, a feature which makes the code particularly efficient when applied to N-body systems characterized with a large dynamical range \citep{Pelupessy2012}.

The code handles sub-systems such as binary and multiple systems, fly-bys and small clusters around supermassive black holes, through the use of secular and regularized few-body integrators \citep{Rantala2020}. The code also includes the option of enabling post-Newtonian equations of motion up to order PN3.5 utilizing the formulas given in \cite{Thorne1985} and \cite{Blanchet2006}. In binary systems the PN corrections account for relativistic effects such as periastron precession and radiation-reaction (circularization and shrinking of the orbit) due to the emission of gravitational waves. Another feature which few codes share \citep{Wang_2020} is the ability to efficiently simulate massive systems with arbitrary fraction of primordial binary systems.

\bifrost{} makes use of four different criteria for stellar and compact object mergers, which we describe in detail Section \ref{sec:merger_criteria}. Those involve merging events of compact objects, disruption of a star by a CO or stellar collisions. For unbound stars at large distances from the centre-of-mass of the system, we employ a user-defined threshold at which the particles are removed from the simulation. For a typical star cluster, stars and COs found beyond $r_\mathrm{esc} \sim 100\; r_\mathrm{h}$ are not considered cluster members anymore, since at such distances the gravitational pull from the parent cluster would be too weak against the galactic tides. We note that the simulations presented in this study do not for simplicity include any tidal potential. The main user-defined code parameters are listed in Table \ref{tab: overview}. Overall the parameters \bifrost{} used are similar to used in previous \bifrost{}studies (e.g. \citealt{Rantala2023,Rizzuto_2023}).

\begin{table}
	\begin{tabular}{lcc} 
		\hline
        \hline
		\bifrost{} user-given parameter & symbol & value\\
		\hline
		forward integrator time-step factor & $\eta_\mathrm{{ff}},\eta_\mathrm{{fb}},\eta_\mathrm{{\nabla}}$ & 0.2 \\
		subsystem neighbour radius & $r_\mathrm{{rgb}}$ & 10 $\mathrm{mpc}$ \\
		regularization GBS tolerance & $\eta_\mathrm{{GBS}}$ & $10^{-7}$ \\
        regularization GBS end-time tolerance & $\eta_\mathrm{{endtime}}$ & $0.01$ \\
        regularization highest PN order &  & PN3.5\\
        secular integration threshold & $N_{\mathrm{orb,sec}}$ & 0 \\
        secular highest PN order & & PN2.5  \\
		\hline
	\end{tabular}
	\caption{The \bifrost{} user-given parameters relevant for simulations of this study. The parameter definitions correspond to the ones in \citet{Rantala2023}.}
    \label{tab: overview}
\end{table}

\subsection{Orbital evolution of binary systems}\label{section: PN}
Binary systems that are strongly perturbed or post-Newtonian are treated in \bifrost{} through algorithmic regularization. We make use of secular binary integration when a binary is sufficiently isolated (no other stars or COs within $10$ mpc) and the number of binary orbits per time-step $\epsilon$, $N_{\mathrm{orb}}=P_{\mathrm{bin}}/\epsilon$ is larger than the user-defined threshold $N_{\mathrm{orb,sec}}$, where $P_{\mathrm{bin}}$ is the orbital period of the binary. We use $N_{\mathrm{orb,sec}}=0$ throughout the study. The initial conditions of the simulations presented here do not contain a binary population. We only form binary systems dynamically. 

The equations of motion responsible for the secular evolution are characterised by the semi-major axis $a$, orbital eccentricity $e$ and the argument of periapsis $\omega$, taking into account leading-order relativistic effects which originate from the post-Newtonian PN1.0, PN2.0 and PN2.5. The advance of the orbital periapsis is due to PN1.0 and PN2.0 given by
\begin{equation}\label{eq: secular_omega}
\begin{split}
\left\langle \derfrac{\omega}{t} \right\rangle_\mathrm{sec} &= \frac{6 \pi G}{c^2 P}\frac{M}{a(1-e^2)} + \frac{3(18+e^2) \pi}{2c^4 P} \left[ \frac{GM}{a(1-e^2)} \right]^2.
\end{split}
\end{equation}
The PN2.5 term refers to gravitational wave radiation reaction due to energy and angular momentum losses, leading to the circularization and shrinkage of the orbit described by the rate of change of binary semi-major axis and eccentricity \citep{Peters1964} as

\begin{equation}\label{eq: secular_a_ecc}
\begin{split}
\left\langle \derfrac{a}{t} \right\rangle_\mathrm{sec} &= -\frac{64}{5} \frac{\beta(m_\mathrm{1},m_\mathrm{2})}{a^3} F(e)\\
\left\langle \derfrac{e}{t} \right\rangle_\mathrm{sec} &= -\frac{304}{15} \frac{\beta(m_\mathrm{1},m_\mathrm{2})}{a^4} e G(e)
\end{split}
\end{equation}
in which the auxiliary functions $\beta(m_\mathrm{1},m_\mathrm{2})$, $F(e)$ and $G(e)$ are defined as
\begin{equation}\label{eq: gw_aux_functions}
\begin{split}
\beta(m_\mathrm{1},m_\mathrm{2}) &= \frac{G^3 m_\mathrm{1} m_\mathrm{2} (m_\mathrm{1}+m_\mathrm{2})}{c^5}\\
F(e) &= \frac{1+\frac{73}{24} e^2 + \frac{37}{96}e^4}{\left( 1-e^2\right)^{7/2}}\\
G(e) &= \frac{1+\frac{121}{304} e^2}{\left( 1-e^2\right)^{5/2}}.
\end{split}
\end{equation}
The secular equations of motion are integrated with the use of a second-order leapfrog integrator, as described in detail in \cite{Rantala2023}.
\subsection{Merger Criteria}\label{sec:merger_criteria}
During a simulation run two objects can merge if any of the four merger criteria is satisfied. Stars can become gravitationally unbound due to strong encounters with the other cluster members. The type of merging pairs can either be compact objects (white dwarfs, neutron stars, BHs) , tidal disruptions (BH with star) and stellar mergers (stars with stars). The first criterion for a compact binary merger depends on the  gravitational-wave driven inspiral timescale $\mathrm{t_{gw}}$. If $\mathrm{t_{gw}}$ is shorter that the binary's current time-step in the time-step hierarchy, then the particles are merged. Compact objects also merge if the radius of the (relativistic) innermost stable circular orbit ($R_{\mathrm{ISCO}}$) is larger than their mutual separation. Finally, stars can be tidally disrupted by a compact object (TDE) if the pericentre distance becomes smaller than the tidal disruption radius or in case of a direct collision between two stellar particles, i.e. when their radii overlap.

A common approach for detecting compact object mergers \citep[e.g.][]{Rizzuto_2020,Arca_Sedda_2021,2022MNRAS.512..884R,2023MNRAS.526..429A} is to compute the timescale $\tau_\mathrm{gw}$ related to the time $\tau_\mathrm{gw}$ a compact binary needs to reach coalescence. This timescale can be evaluated through the integral expression \citep{Peters1964}

\begin{equation}
\tau_\mathrm{gw} = \frac{15}{304} \frac{a_\mathrm{0}^4}{\beta(m1,m2)} \frac{1}{g^4(e_\mathrm{0})} \int_\mathrm{0}^\mathrm{e_\mathrm{0}} \frac{g^4(e)(1-e^2)^{5/2}}{e \left( 1+\frac{121}{304} e^2\right)} de ,
\label{eq:t_gw}
\end{equation}

where $a_0$ and $e_0$ are the initial semi-major axis and eccentricity of the binary, $\beta(m_\mathrm{1},m_\mathrm{2})$ is the function defined in Eq. \eqref{eq: gw_aux_functions} and $g(e)$ is defined \citep{Maggiore2007} by
\begin{equation}
g(e) = \frac{e^{12/19}}{1-e^2} \left(1 + \frac{121}{304} e^2 \right)^{870/2299}.
\end{equation}

The expression for $\tau_\mathrm{gw}$ for $e=0$ , i.e. circular orbits, is then
\begin{equation}
\tau_\mathrm{gw} = \frac{5}{256} \frac{a_\mathrm{0}^4}{\beta(m_\mathrm{1},m_\mathrm{2})}.
\end{equation}

We tabulate the values of this integral by evaluating it separately (before the actual runs) for a large number of $e_0$, in order to avoid computing it every time-step. Then by interpolating for those table values, we have an estimate for $\tau_\mathrm{gw}$ throughout the simulation. However, extra caution should be taken when using the $\tau_\mathrm{gw}$  criterion due to the strong dependence of eccentricity in the computed integral value. Even tiny fluctuations of eccentricity (e.g. from weak 3- body and fly-by encounters) would result in significant underestimation of $\tau_\mathrm{gw}$ and false merger identification.

The second criterion for merging compact objects is based on the innermost stable circular orbit (ISCO) with radius

\begin{equation}
r_\mathrm{isco} = \frac{6GM_\bullet}{c^2} = 3 R_\mathrm{sch}, 
\end{equation}

where $c$ is the speed of light and $R_\mathrm{sch}$ the Schwarzschild radius of the black hole. Two COs are then merged if their mutual distance is lower than $r_\mathrm{isco}$.

If one of the particles is a compact object and the other one is a star, then the latter could be disrupted. This happens when \citep{1988Natur.333..523R,Kochanek1992}
\begin{equation}
r_\mathrm{peri} < r_\mathrm{tde} = 1.3 \left( \frac{m_\star + M_\bullet}{m_\star} \right)^{1/3} r_\star,
\end{equation}
where $m_\star$ and $r_\star$ is the mass and radius of the star and $M_\bullet$ is the mass of the compact object. Two stars are assumed to merge when they overlap i.e., when their distance is shorter than the sum of their radii
\begin{equation}
r < r_\mathrm{overlap} = r_{\star,\mathrm{1}} + r_{\star,\mathrm{2}}.
\end{equation}

\subsection{Initial Conditions and Simulations}

We generate initial conditions (ICs) for pairs of two identical star clusters of $N=64000$ single stars without primordial binaries and a total mass of $M_{\star} = 2.7 \times 10^4 M_{\odot}$,  each using the \mcluster{} code \citep{Kuepper_2011}. The single star cluster model follows a King density profile \citep{King_1966} with a fixed half-mass radius of $R_{h}=1 \: \mathrm{pc}$ and central potential parameter related to the compactness of the cluster $W_{0}=5$, which represents a moderately compact cluster. The age of the clusters is $t_{\mathrm{age}}= 1 \: \mathrm{Gyr}$ and the stellar masses are sampled from a Kroupa initial-mass-function (IMF) \citep{Kroupa_2001} as zero-age main sequence stars in the range of $0.08 M_{\odot}$ up to $100 M_{\odot}$. The age of the cluster has been selected such that stellar evolution processes do not contribute to the overall evolution of the system, allowing us to focus on the long-term dynamical evolution. The mass range of stellar particles in the simulations is $0.08 \leq m_{\star} \lesssim 1.89 \: \mathrm{M_{\odot}}$, $0.88 \leq m_{\mathrm{WD}} \lesssim 1.37 \: \mathrm{M_{\odot}}$ for WDs, $1.1 \leq m_{\mathrm{NS}} \lesssim 1.91 \: \mathrm{M_{\odot}}$ for NSs and $5.6 \leq m_{\mathrm{BH}} \lesssim 40.5 \: \mathrm{M_{\odot}}$ for stellar-mass BHs. From the initial conditions, we have initially $N_{\mathrm{BH}}=250$, $N_{\mathrm{NS}}=944$ and $N_{\mathrm{WDs}}=4100$. The number of retained NSs in the initial conditions is related to the way natal kicks for supernova remnants are assigned in \cite{Kuepper_2011} , i.e., when the recoil velocities are low so the remnant cannot escape the cluster. With the use and calibration of pulsar proper motions \citep{Hobbs_2005, Kapil_2023} the recoil kicks reach high velocities resulting in small fraction (only a few) of retained NSs in the initial system. Examples of a more sophisticated prescription for the number of retained NSs in the ICs can be found in \cite{Banerjee_2020}. The NS population in our simulations does not significantly contribute to the overall ejection demographics, but could potentially lead to false rates of escaping NSs and the formation of a larger number of NS binaries.

We use three different initially bound orbits for the merging star clusters. Each orbit has the same fixed semi-major axis $a_{\mathrm{semi}}=2 \: \mathrm{pc}$ (and thus a fixed orbital energy, since $E = -G M_1 M_2/2a_{\mathrm{semi}}$) and three different values of eccentricity $e=[0.05,0.5,0.99]$. The second cluster is located at the apocentre of the orbit with $r_{\mathrm{apo}}=[1.9,1.0,0.02] \: \mathrm{pc}$ for the three orbits. For each of the merger initial setups, we explore three scenarios: neither of the clusters host an MBH, the primary cluster host an MBH of $M_{\bullet}=M_1=1000 \: M_{\odot}$ and finally when both clusters host an MBH of $M_1=M_2=500 \: M_{\odot}$. Furthermore, we run three additional simulations where we consider a star cluster with $N=128000$, $M_{\star} = 5.4 \times 10^4 \: M_{\odot}$ and $R_{h}=1 \: \mathrm{pc}$ without a prior merger. For these simulations we either place a single, a binary with $a_{\mathrm{b}}= 0.01 \: \mathrm{pc}$ and $e_{\mathrm{b}}= \: 0.5$, or no MBH at all. We follow the evolution of the merger remnants or the single clusters up to $t=100 \: \mathrm{Myr}$. An overview of the merger progenitors and their orbital elements is presented on table \ref{tab:ICs} . 

\begin{table}
\begin{tabular}{l|cccccc} 
\hline
\hline
simulation & $M_{\mathrm{\bullet,1}}$ & $M_{\mathrm{\bullet,2}}$ & $e$ & $r_{\mathrm{peri}}$ & $v_{\mathrm{peri}}$ & $v_{\mathrm{apo}}$ \\
                   & $[\mathrm{M_{\odot}}]$ & $[\mathrm{M_{\odot}}]$ &  & $[\mathrm{pc}]$ & $[\mathrm{km/s}]$ & $[\mathrm{km/s}]$ \\
\hline
          2MBH\textunderscore e099 & 500 & 500 & 0.99 & 0.02 & 151.89 & 0.76  \\
          2MBH\textunderscore e05  & 500 & 500 & 0.50 & 1.0  & 18.65  & 6.22  \\
          2MBH\textunderscore e05  & 500 & 500 & 0.05 & 1.9  & 11.32  & 10.24 \\
          1MBH\textunderscore e099 & 1000 & -  & 0.99 & 0.02 & 151.89 & 0.76  \\
          1MBH\textunderscore e05  & 1000 & -  & 0.50 & 1.0  & 18.65  & 6.22  \\
          1MBH\textunderscore e005 & 1000 & -  & 0.05 & 1.9  & 11.32  & 10.24 \\
          noMBH\textunderscore e099 &   -  & -  & 0.99 & 0.02 & 150.48 & 0.75  \\
          noMBH\textunderscore e05  &   -  & -  & 0.50 & 1.0  & 18.47  & 6.16  \\
          noMBH\textunderscore e005 &   -  & -  & 0.05 & 1.9 & 11.21  & 10.15 \\
          2MBH\textunderscore iso  & 500  & 500 &  -  & - & - & - \\
          1MBH\textunderscore iso  & 1000 &  -  &  -  & - & - & - \\
          noMBH\textunderscore iso  &  -   &  -  &  -  & - & - & - \\
\end{tabular}%
\caption{List of cluster merger simulations with two (2MBH), one (1MBH) or no MBH (noMBH) with three cluster orbital eccentricities $e$ (fourth column). The semi-major axis of all merger obits is 2 pc. We also give the masses ($M_{\mathrm{\bullet,1}}, M_{\mathrm{\bullet,2}}$) of the MBHs as well as the pericentre distance and velocity ($r_{\mathrm{peri}}$, $v_{\mathrm{peri}}$ and the apocentre velocity ($v_{\mathrm{apo}}$) of the initial star cluster orbits. As the star clusters a not point masses they will not follow this initial orbit but rather merge due to dynamical friction. The bottom three simulations ($\_iso$) correspond to the isolated setups with no, one and two MBHs with an eccentricity of $e = 0.5$.}
    \label{tab:ICs}
\end{table}

In the next sections, we present the main outcomes of our stellar cluster merger simulations. We discuss how the presence of a single or a binary MBH affects the evolution of the merger remnants, their kinematic and structure properties as well as the demographics of the escaping stellar and compact object populations, which is the main focus of this work.  An example outlook (here for the $e=0.05$ cluster orbits) of the simulations at different times ($ t = 0.5,1,15 \: \mathrm{Myr}$), is shown in Fig. \ref{fig:initial_final_cl}. In the figure we show the surface stellar density and highlight the velocity vector $\vec{V}$ field of unbound stars and COs.

If an MBH is present in both of the merging clusters, a massive black hole binary (MBHB) forms. The binary orbit keeps shrinking (monotonic decrease of $a_{\mathrm{b}}$) until the end of the simulation, due to interactions with the background stars and COs. We describe the evolution of the MBHBs in detail in Section \ref{sec:MBHB_evol}.

\begin{figure}
	\includegraphics[width=0.9\columnwidth]{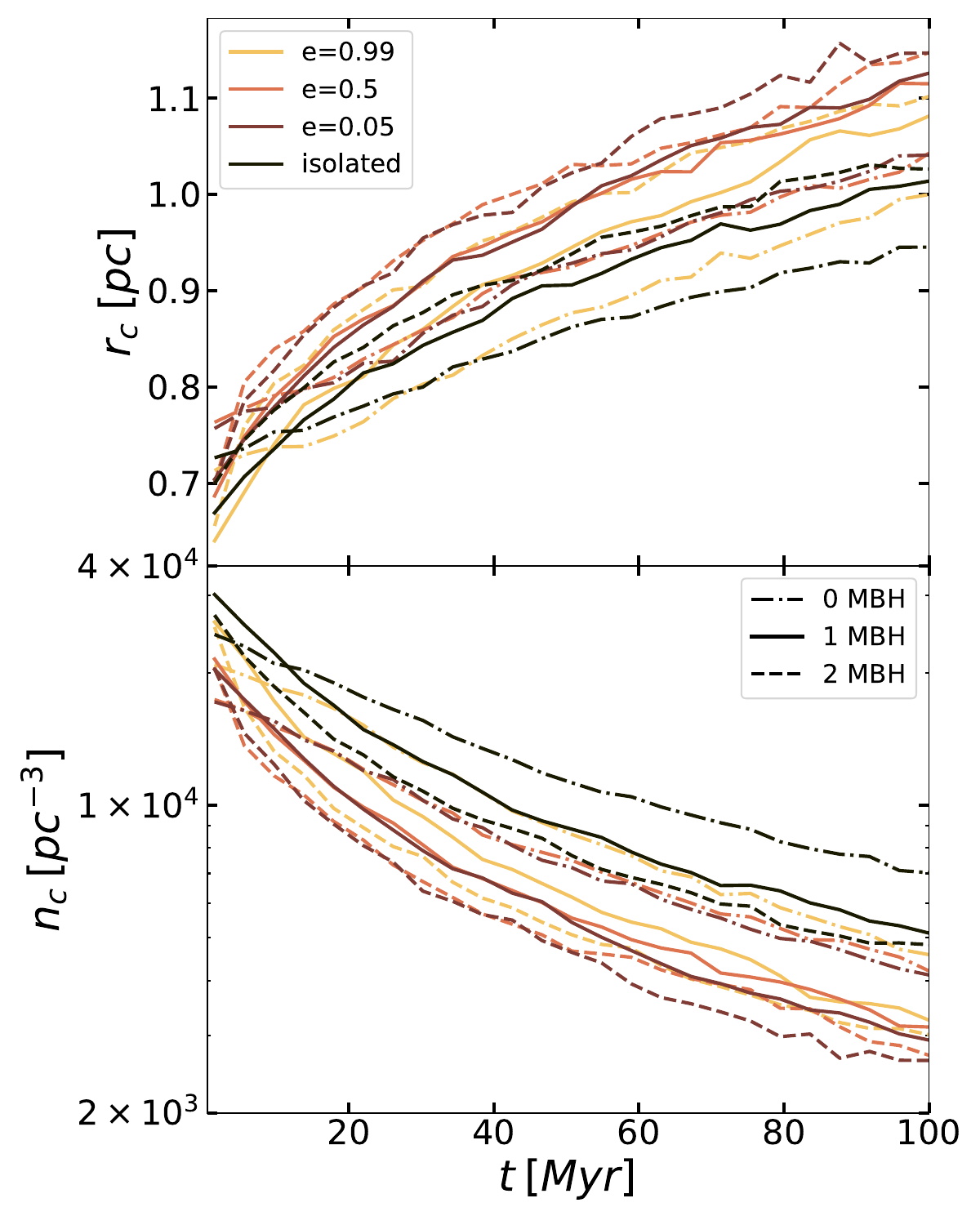}
    \caption{Evolution of the core for the three cases. Top panel: Core radius $r_{c}$. Bottom panel: Number density inside $r_{c}$. The expansion rate (decreasing number density) is higher for remnants with more MBHs and lower values of eccentricity.}
    \label{fig:core}
\end{figure}

\section{Properties of the star clusters}
\label{sec:clusters}
Adopting the definition by \cite{Trenti_2009}, the core size or \textit{core radius} of a star cluster following a King density profile is given by \cite{Casertano_1985},

\begin{equation}
r_{\mathrm{c}}=\sqrt{\frac{\sum_i \rho_i^2\left|\mathbf{r}_i-\mathbf{r}_{\mathrm{d}}\right|^2}{\sum_i \rho_i^2}},
\label{eq:core_rad}
\end{equation}

where $\rho_i$ is the local density around a star $i$ defined as

\begin{equation}
\rho_i=\frac{\sum_{j=1}^5 m_j}{\frac{4}{3} \pi r_n^3}
\end{equation}

and $r_n$ is the distance to the n-th nearest neighbor
$i$, while $\mathbf{r}_{\mathrm{d}}$ is the position of the density centre:

\begin{equation}
\mathbf{r}_{\mathrm{d}}=\frac{\sum_i \rho_i \mathbf{r}_i}{\sum_i \rho_i}.
\end{equation}

When the cores of the two progenitor clusters overlap, i.e., when their relative distance $d_{12}$ becomes shorter than $r_c$, we assume that the clusters have merged. The evolution of the core radii and the respective enclosed number density is shown in Fig. \ref{fig:core}. From Fig. \ref{fig:core} we see that the presence of a single or binary MBH in the remnants leads to more rapid expansion due to a higher mass-loss rate. Essentially, particles interacting with the MBHs are radially scattered outwards, leading to constant decrease of the number density inside the core (bottom panel in Fig. \ref{fig:core}) which indicates mass-loss inside $r_c$.


Since the core radius $r_c$ is not exactly the same (ranging between $0.6\mathrm{pc}<r_c<0.8\mathrm{pc}$) for the various progenitor clusters, we define a threshold distance such that when $d_{\mathrm{12}}<d_{\mathrm{crit}}=2 \mathrm{pc}$, the two clusters have merged. The timescale of this process depends on the orbital eccentricity of the merger and on the number of MBHs in the system. Specifically for remnants with two MBHs, this happens at $t=1 \: \mathrm{Myr}$ while for the other cases ranges between $t=1.25 \: \mathrm{Myr}$ and $t=2 \: \mathrm{Myr}$. 

\begin{figure*}
	\includegraphics[width=0.8\textwidth]{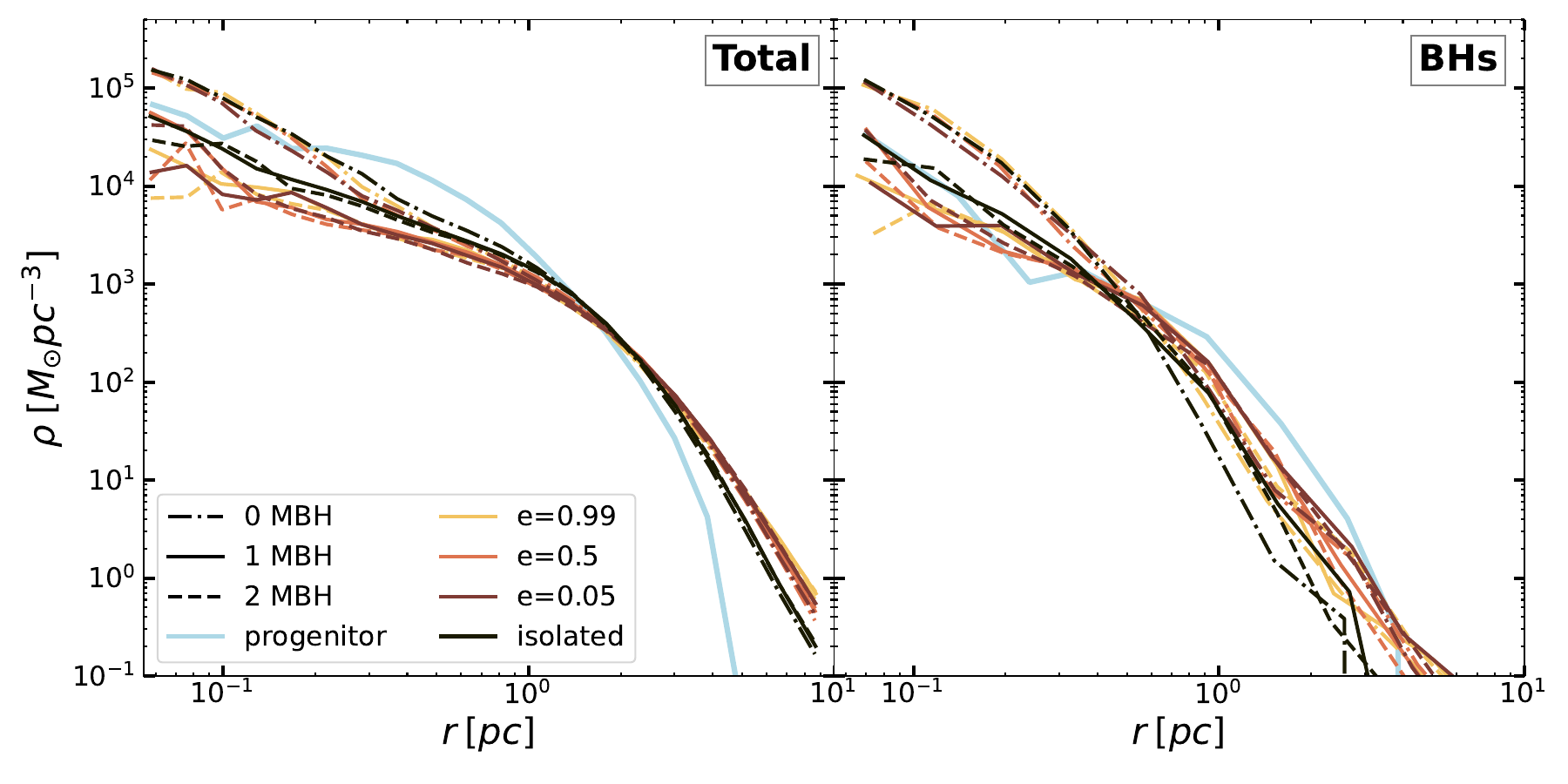}
    \caption{Final density $(t=100 \: \mathrm{Myr})$ profiles of the merger remnants. Left panel: density profiles including all cluster members. Right panel: density profiles of the stellar BH sub-systems. Remnants hosting a single or binary MBH result in lower density in the central regions. For BH sub-systems in the absence of MBHs to provide a counteracting source of contraction, the corresponding density profiles become more cuspy.}
    \label{fig:density_profiles}
\end{figure*}

The density evolution for $t>2 \: \mathrm{Myr}$ until the end of the run, is only moderately affected by the merger process and its mainly driven by the presence of the single or binary MBHs. Figure \ref{fig:density_profiles} shows the final (at $t = 100\mathrm{Myr}$) density profiles of the merger remnants and that of the stellar BHs only. Overall we observe an increase of the central density, which inversely scales to the presence and number of MBHs in the remnant, with an even more clear signature for the sub-system of BHs (right panel in Fig. \ref{fig:density_profiles}). The binary transfers its orbital energy

\begin{equation}
E=-\frac{G M_1 M_2}{2 a_{\mathrm{b}}},
\end{equation}

to the surrounding stars either through dynamical friction (DF) \citep{Begelman_1980Natur, Varisco_2021} or three-body encounters \cite{Hills_1991} accompanied by stellar ejections which carry out energy and angular momentum, leading to a decrease of stellar density around the MBHB. This process, known as \textit{core scouring}, is likely responsible for the presence and formation of flat central cores in luminous galaxies (e.g. \citealt{Rantala_2018}), and can likely affect the central properties of merged star cluster remnants as well. The characteristic timescale over which the energy is extracted from the binary is given by \cite{Merritt_2013_book},

\begin{equation}
T_{\mathrm{E}}=\left|\frac{1}{E} \frac{\mathrm{~d} E}{\mathrm{~d} t}\right|^{-1} \approx \frac{\sigma^3}{C G^2 M_2 \rho} ,
\end{equation}

where $\sigma$ and $\rho$ are the three-dimensional velocity dispersion and density of the stellar background and $C$ is a constant that depends on the energy transfer mechanism\footnote{$C \approx 10$ for DF or 3-body encounters.}. In our simulation models we have $T_{\mathrm{E}} = 1 - 3.7 \: \mathrm{Myr}$, which is 
consistent with the time needed $(\sim 1 \: \mathrm{Myr})$ for the respective progenitor clusters to merge.

\subsection{Star cluster evolution}
\label{sec:remnant_evol}
\begin{figure*}
	\includegraphics[width=0.49\textwidth, keepaspectratio]{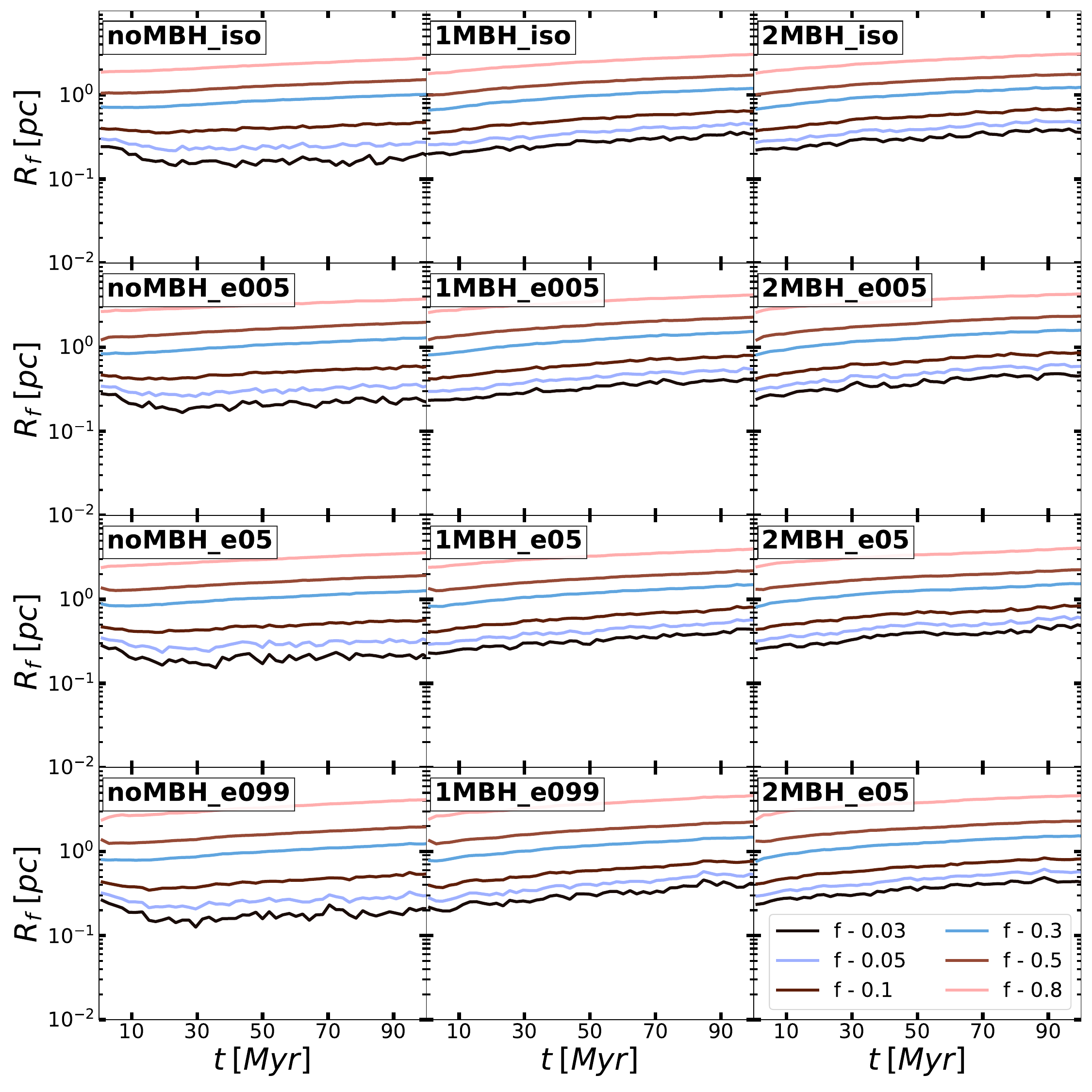}
 \includegraphics[width=0.49 \textwidth, keepaspectratio]{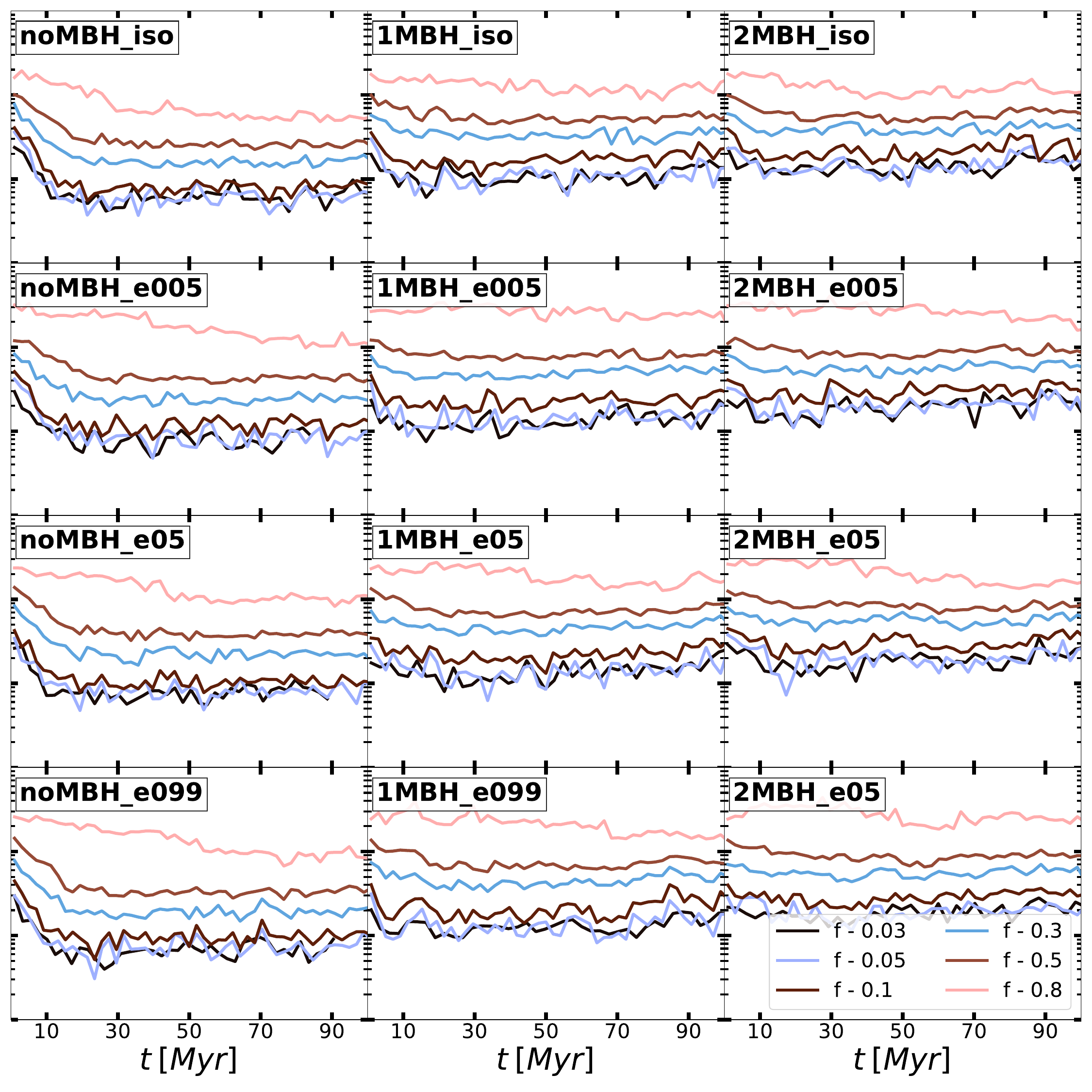}
    \caption{Lagrangian radii for the $3-80 \%$ enclosed mass of the entire system (top plot) and for the respective BH subsystems only (bottom plot). Columns from left to right refer to different number of MBHs, while rows to different orbital eccentricities of the merger orbits. The top row in each panels shows the isolated runs. An initial central collapse phase is clearly seen in the runs without MBHs (left columns).}
    \label{fig:lagrange_rad}
\end{figure*}

A bound self-gravitating system admits negative heat capacity $C_{\mathrm{V}}$, allowing the release of energy from the centre to the outer parts, leading to the formation of a contracting core with radius $r_c$ (Eq. \ref{eq:core_rad}) and an expanding halo \citep{Lynden_Bell_1968, Lynden_Bell_1999}. 

In the absence of an additional energy source in the centre to halt the contraction the core will eventually collapse, a process called the \textit{gravothermal catastrophe} \citep{Antonov_1960,Antonov_1961,Antonov_1962}. The core undergoes a series of contraction-expansion phase, known as \textit{gravothermal oscillations} (\citealt{Sugimoto_1983, Heggie_1993}). For a single and equal mass component system in isolation, the core collapse would happen at about $15t_{\mathrm{rh}}$ \citep{Cohn_1980}, where $t_{\mathrm{rh}}$ is the half-mass relaxation time \citep{Spitzer_1987}.

\begin{equation}
t_{\mathrm{rh}}=\frac{0.138 N^{1 / 2}}{\ln \Lambda}\left(\frac{R_{\mathrm{h}}^3}{G \bar{m}}\right)^{1 / 2}.
\end{equation}

Here $N$ is the star/particle number of the system, $R_h$ and $\bar{m}$ the half-mass radius and average stellar mass and finally $\Lambda=\gamma N$ is the Coulomb logarithm. In our simulations we sample the stellar masses from a Kroupa IMF\footnote{For a single-mass system $\gamma=0.11$ \citep{Giersz_1994}.}, which corresponds to $\gamma=0.02$ \citep{Giersz_1996} for multi-mass clusters. In realistic (multi-mass) star clusters, higher-mass stars have lower velocity dispersion than the low-mass ones, driving the system towards energy equipartition, i.e., evenly distributing the kinetic energy of the system \citep{Spitzer_1969,Spitzer_1987}. The massive stars segregate to the centre by transferring their kinetic energy to low mass stars, thus expanding the orbits of the latter. The segregation time \citep{Spitzer_1971,Portegies_Zwart_2004} is defined as

\begin{equation}
t_{\mathrm{s}}=\frac{\bar{m}}{m_{\max }} \frac{0.138 N}{\ln \left(0.11 M_{\mathrm{cl}} / m_{\max }\right)}\left(\frac{R_{\mathrm{h}}^3}{G M_{\mathrm{cl}}}\right)^{1 / 2},
\label{eq:t_seg}
\end{equation}

where $M_{\mathrm{cl}}$ is the total mass of the cluster and $m_{\mathrm{max}}$ is the mass of the most massive object. Since we don't assume any degree of primordial mass segregation in our initial conditions, we compute the segregation time of the merger remnants at $t= 2 \: \mathrm{Myr}$, when all the progenitors have merged. Using Eq. \ref{eq:t_seg} for the remnant clusters without an MBH, we find $t_{\mathrm{s}} \sim 8-10 \: \mathrm{Myr}$. In  timescales of the order of $\sim t_{\mathrm{s}}$ compact objects, i.e., the most massive particles sink towards the centre of the remnant \citep{Baumgardt_2003}. 

The evolution of the radii enclosing $3-80\%$ of the cumulative total mass (Lagrangian radii) of all stellar and compact objects for all simulations is shown in the top panel of Fig. \ref{fig:lagrange_rad}. The left, middle and right columns show the systems with no, single, or binary MBHs. For the isolated cluster the centre of the system without MBH (left) undergoes core contraction on a timescale of $\sim$ 20 - 30 Myr while the half-mass radius expands. The systems with single and binary MBH (middle and right) show no core collapse as rapidly forming binaries with the central MBH prevent this \citep[see e.g.][and references therein]{Rizzuto_2023}. Stars and compact objects bound to MBHs interact, heating up the core and lead to its mild but constant expansion. The evolutionary behaviour is very similar for cluster merger remnants (second to bottom row). For mergers without a MBH the one with the highest eccentricity (bottom left panel) shows most core contraction. For systems with one or two MBHs the eccentricities of the merger orbits do not change the structural evolution.  

In the bottom panel of Fig. \ref{fig:lagrange_rad} we show the Lagrangian radii evolution of the respective BH subsystems only. Initially, the general behaviour is the same as for the entire cluster (left panels in Fig. \ref{fig:lagrange_rad}). However, the BH subsystem contracts faster on a $\sim$ 10 Myr timescale which is the expected mass segregation timescale (see Eq. \ref{eq:t_seg}) in all cases without MBHs (left column). However, the presence of just one MBH does not prevent the collapse in for the BH subsystem. In addition the BHs form a central, stable subcluster which is not expanding like the entire cluster population. The dynamical interactions in this central BH cluster provide a continuous source of energy for ejections of stars and compact objects.

The sphere around an MBH of mass $M_{\bullet}$ where its presence dominates the gravitational potential of the system is called \textit{sphere of influence} (SOI) . The radius of this sphere $r_{\mathrm{SOI}}$ is then given by \citep{Merritt_2013_book}

\begin{equation}
\label{eq:rSOI}
    M_{\star}(r < r_\mathrm{SOI}) = 2 M_{\bullet},
\end{equation}

where $M_{\star}$ refers to the stellar mass.

\begin{figure}
	\includegraphics[width=0.9\columnwidth]{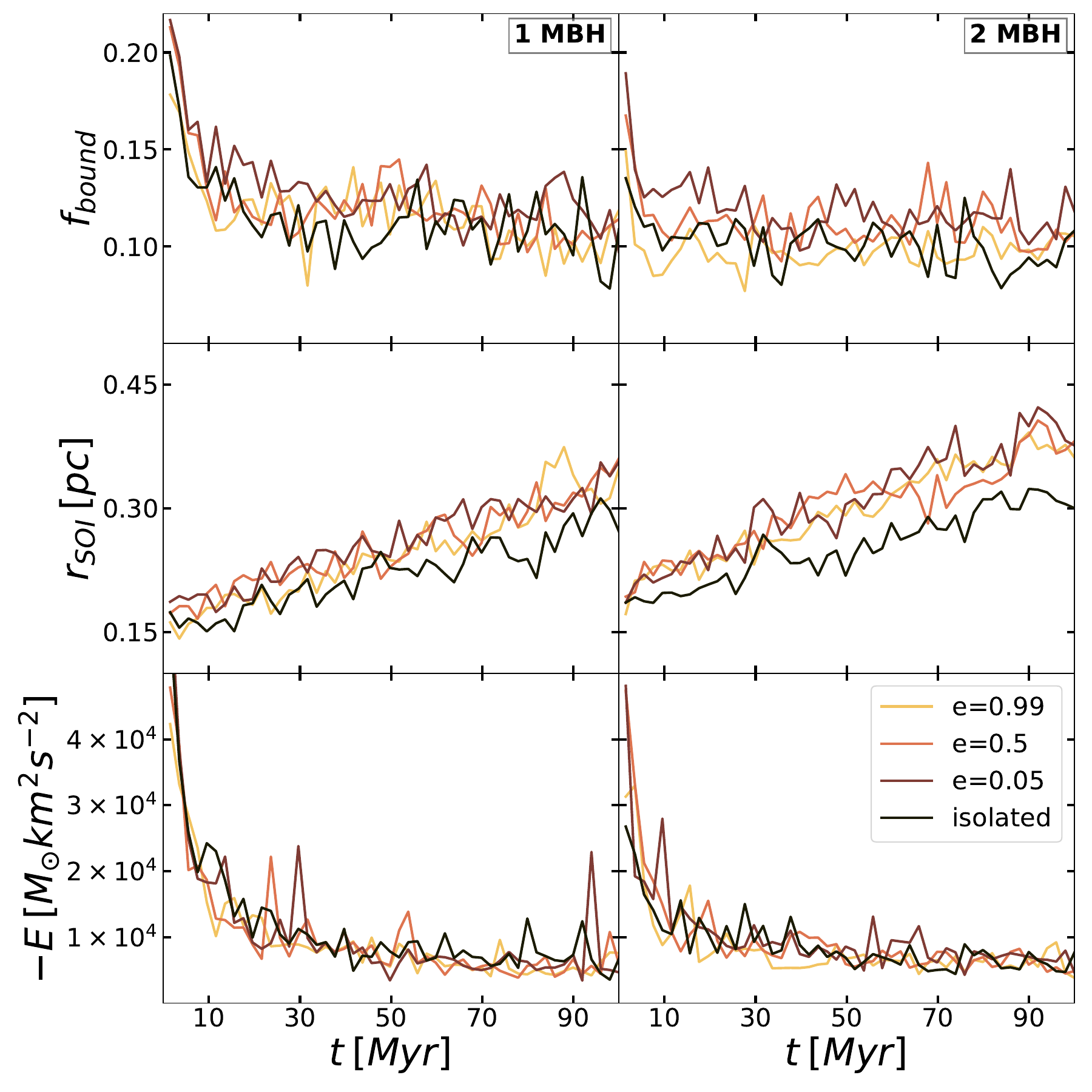}
    \caption{Top: fraction of bound stars $f_{\mathrm{bound}}$ inside the sphere of influence $r_{\mathrm{SOI}}$ of the single/binary MBHs. Middle: MBH radius of influence. Bottom: Energy inside $r_{\mathrm{SOI}}$. Both Single and binary MBHs initially have a high fraction of bound stars, which is decreased over time due to encounters inside $r_{\mathrm{SOI}}$, leading to energy losses elevating its growth. The orbital energy released from the hardening of the MBH binary also contributes to the expansion. }
    \label{fig:bound_rSOI}
\end{figure}

The interactions of stars within $r_{\mathrm{SOI}}$ with the MBH/MBHB can significantly effect their growth through TDEs \cite{Rizzuto_2023} or GW-merger events, but also play an important role on the evolution of the remnant cores \citep[e.g.][and references therein]{Aros_2020}{}{}. Figure \ref{fig:bound_rSOI} shows the fraction of bound stars to the single or binary MBH, i.e., stars with negative orbital energy $E=\frac{m_* v^2}{2}-\frac{G m_* M_{\mathrm{BH}}}{r}<0$, where $m_{*}$ is the mass of the individual star and $r,v$ are the relative distance and velocity w.r.t. the MBH or the centre of mass of the MBH binary. After the cluster mergers the bound fractions are $f_\mathrm{bound}\sim0.14$--$0.22$ in our simulations. The MBH case behaves in a qualitatively similar manner as the binary MBH setup: the bound fractions reach values of $f_\mathrm{bound}\sim0.1$ in both cases. This reflects the similar overall evolution of central cluster properties such as stellar densities and velocity dispersions. We however note that the bound fraction decreases somewhat more rapidly in the MBHB setups. For clusters with a single MBH we find similar behavior to that of the $M_{\bullet}=2000 \: M_{\odot}$ case in \citep{Rizzuto_2023}. We note though, that their clusters are more dense and that stellar-mass BHs were not present in their ICs. The remnants with an MBH binary, maintain approximately the same fraction of bound stars until the end of the simulation. The bounds stars in the vicinity of the MBH or the MBHB provide enough energy supply to support the expansion of the core (Fig. \ref{fig:core}) and the remnant itself. Moreover, encounters with the MBHB lead to 3- and 4- body encounters (Section \ref{sec:MBHB_encounters}), capable of kicking stars and COs in the outer parts of the cluster. The lower panel of Figure \ref{fig:bound_rSOI} depicts the total energy loss from the bound cloud of stars inside $r_\mathrm{SOI}$ leading to its growth. Those encounters drain orbital energy from the MBH binaries, leading to the shrinking of their orbits (decrease of semi-major axis) . This hardening process is further discussed in Section \ref{sec:MBHB_evol}. The energy inside the cloud in the form of dynamical heating combined with that from the BH sub-system \citep{Mackey_2008} leads to the expansion of the remnant cores and monotonic decrease of the number density (bottom panel in Fig. \ref{fig:core}). Subsequently, the drop of the number density is associated with a decrease in the mass of the bound cloud, i.e., the gravitational potential of the core gradually becomes more shallow (i.e., a drop in the local escape velocity $V_{\mathrm{esc}}$), allowing more particles to escape to the outer parts of the cluster. 

Finally the merger process itself has an effect on the differential energy distribution of the cluster stars. For a bound gravitational system in virial equilibrium $2T+V=0$ (kinetic energy $T$ and potential energy $V$) with $T=-E$ and $V=2E$. The total energy is simply $E=T+V$. During the merger, the system is not in equilibrium which means that although the total energy $E$ remains constant, $T$ and $V$ will oscillate around their values leading to the widening of the differential energy distribution \citep{Hilz_2012}. This results to the most bound particles become even more bound, while weakly bound particles might gain enough energy to escape from the local gravitational potential.

This process, known as \textit{violent relaxation} further contributes to the expansion of the core and the cluster itself, comparing to clusters without any prior merger (dashed lines in \ref{fig:core}) .

\begin{figure*}
  \includegraphics[width=0.8\textwidth]{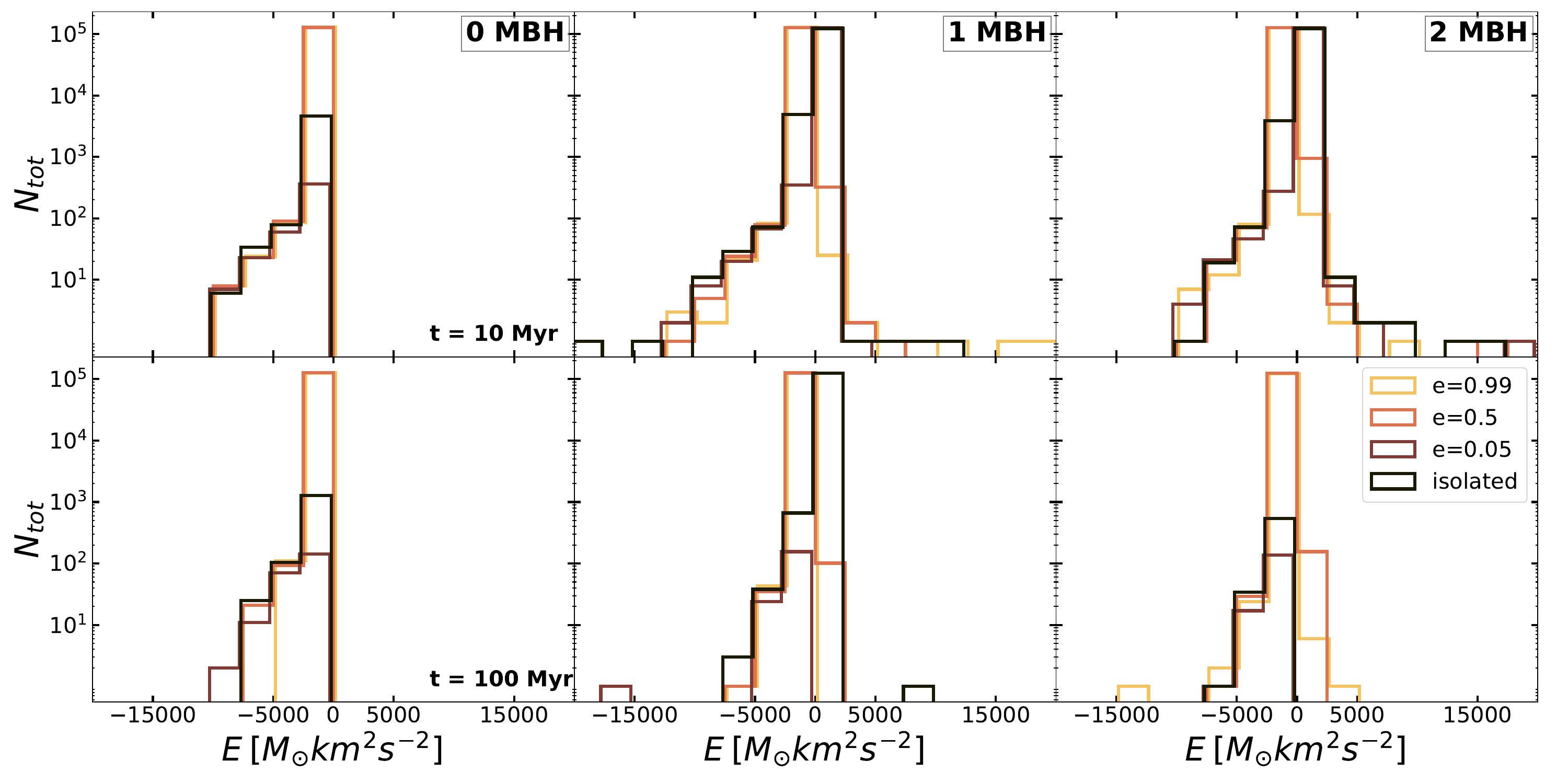}
  \caption{Binding energy distribution for stars and compact objects in all simulated clusters at $t=10$ Myr (top) and $t=100$ Myr (bottom). Once the systems contain at least one MBH (middle and left panels) objects can become unbound (positive energies). In particular at early times ($t=10 Myr$, middle panel) the high energy objects can be clearly seen. At later times many of the early escapers have travelled beyond 100 pc and have been removed from the simulation. There is no clear trend with the shape of the cluster merger orbit or the merging process itself.}
  \label{fig:Edist}
\end{figure*}

\begin{figure*}
  \includegraphics[width=0.8 \textwidth]{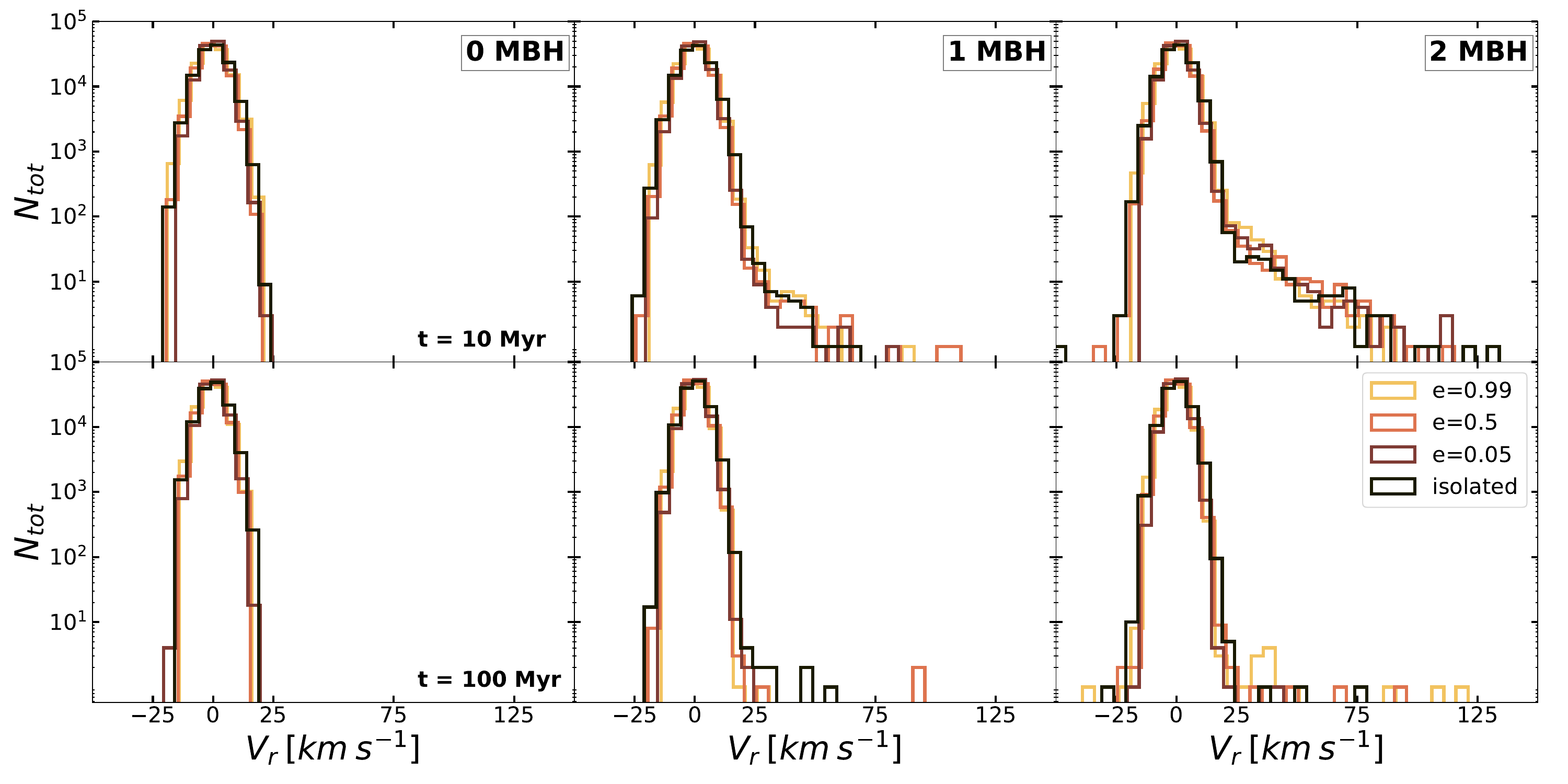}
  \caption{Distribution of the radial velocity of all stars and compact objects in all simulations at $t=10$ Myr (top) and $t=100$ Myr (bottom). The systems without MBHs (left panels) show a distributions with absolute radial velocities $\leq$ 23 km s$^{-1}$. This is the escape velocity from the cluster centres. Once an MBH is present, the distributions extend towards $\sim$ 75 km s$^{-1}$ for one MBH (middle panels) and to values $\gtrsim$ 120 km s$^{-1}$ for binary MBHs (right panels). The unbound part of the distributions in Fig. \ref{fig:Edist} is caused by objects within the high velocity tail.}
  \label{fig:Vrdist}
\end{figure*}

Figure \ref{fig:Edist} shows the particle energy distribution at $t=10$ and $t=100 \: \mathrm{Myr}$. We see that clusters with MBHs hold a large fraction of unbound particles at early times corresponding to potential ejections. It is clear that the presence of MBHs broadens the energy distribution, especially for stars with $E>0$ we observe a persistent tail in the distribution even after $t=100\mathrm{Myr}$. Especially, from the radial velocity $V_{\mathrm{r}}$ distribution (Fig. \ref{fig:Vrdist}), we see that single MBH clusters develop a positive $V_{\mathrm{r}}$ tail which is larger for MBH binaries. This tailed distribution is closely connected to the number and velocity of ejected bodies as we describe in Section \ref{section: ejections}.

\subsection{Star cluster shapes}

In order to determine the inherent three-dimensional structure of the star clusters, we compute the reduced inertia tensor given by given by \cite{Gerhard_1983,Bailin_2005} 

\begin{equation}
\label{eq:inertia_tensor}
    I_{ij} = \sum_k m_k \frac{r_{k,i} r_{k,j}}{r_k^2},
\end{equation}

where $k$ is the total particle number. The eigenvectors and eigenvalues of this tensor correspond to the directions of the principal axes $a,b,c$ (major, intermediate and minor axis respectively, i.e., $a>b>c$) of the remnant. Their ratios can then be used to define the triaxiality parameter $T$ \citep{Jesseit_2005, Binney2008} given by

\begin{equation}
T=\frac{1-(b / a)^2}{1-(c / a)^2} .
\end{equation}

\begin{figure}
	\includegraphics[width=0.9\columnwidth]{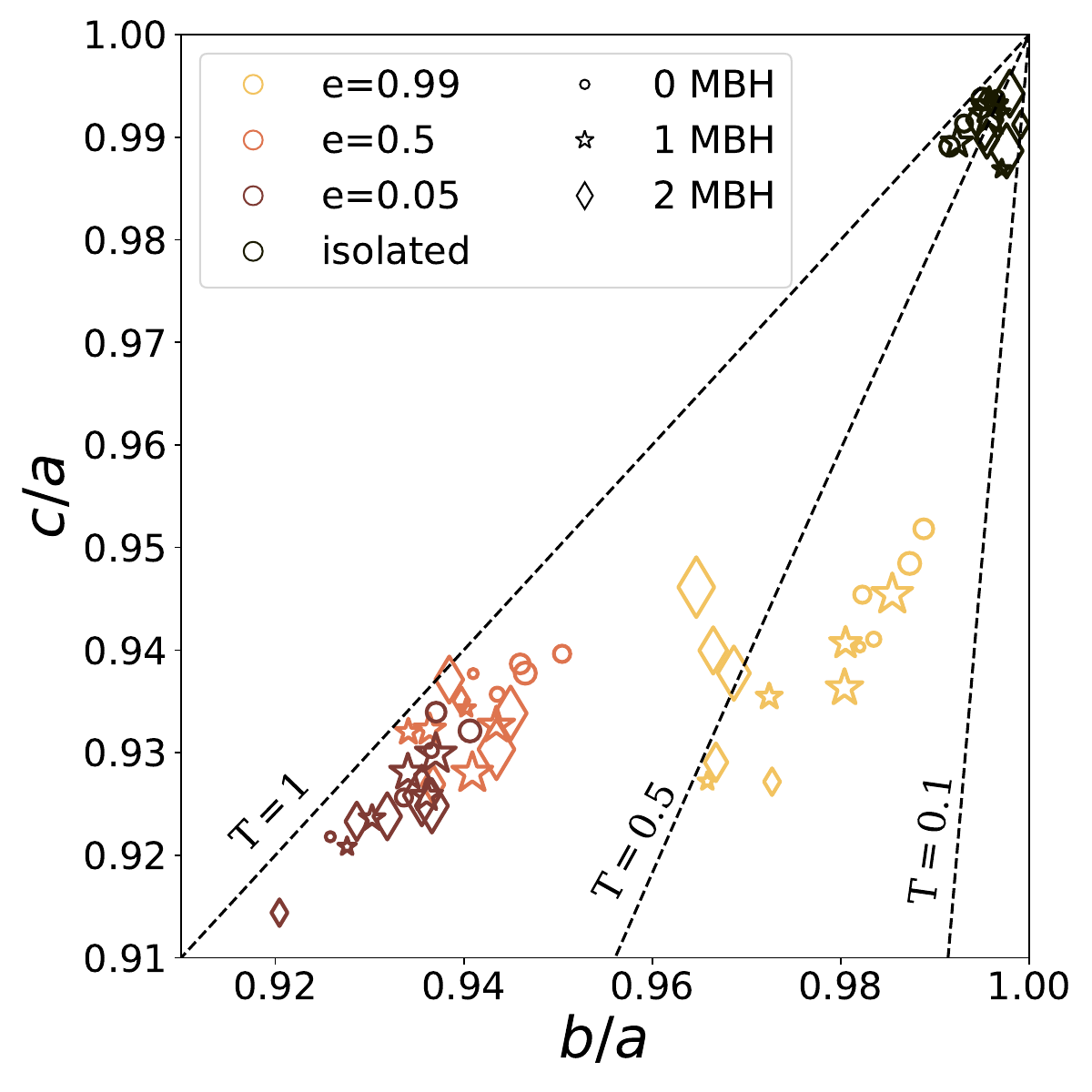}
    \caption{Axis rations of the three principal axes of the moment-of-inertia tensor within the $r_{\mathrm{h}}$. The dashed lines correspond to constant triaxiality $\mathrm{T}$, where $T=1:$ prolate, $T=0.5:$ triaxial, $T=0.1:$ oblate. Different colours here correspond to different values of eccentricity, while markers to the number of MBHs. The size of the markers grows with time, i.e., the smaller one is at $t=0$. }
    \label{fig:triax}
\end{figure}

Fig. \ref{fig:triax} shows the evolution of the principal axes ratios and the triaxiality parameter within the half-mass radius $r_{\mathrm{h}}$, where $T=1$ correspond to prolate, $T=0.5$ to triaxial and $T=0$ to oblate shapes. 

The shape of merger remnants is an indicator of the orbital families that are generated during the merger process (for a detailed analysis on a galaxy merger context, see for example \citealt{Frigo_2021}). In Fig. \ref{fig:triax} we present the time evolution of the remnant shapes. The size of the markers indicates the time-evolution of cluster shapes, where the largest ones correspond to $t=100 \: \mathrm{Myr}$ (end of the simulation). Constant triaxiality $T$ is marked with the three dashed lines. This provides a qualitative way to classify the resulting shapes for the various models based on their evolution and initial merger orbit. First of all, we notice that the clusters without prior merger are initially spherical and isotropic and maintain their shape (black markers in the top-right corner of Fig. \ref{fig:triax}) until the end of the simulation.

The eccentricity of the merging star cluster orbits plays an important role in determining the shape and kinematics of the remnant. Low $e$ orbits carry more angular momentum, since $\mathcal{L}=\mu \sqrt{G M a\left(1-e^2\right)}$ (where $M=M_1+M_2$ and $\mu=M_{1}M_{2}/M$) which is then transferred to the stars. Our merger simulations result in a narrow range of $c/a$ with a clear trend of low ($e=0.05$) and medium eccentricities ($e=0.5$) towards less oblate systems (brown and orange markers in Fig. \ref{fig:triax}). Very eccentric ($e>0.9$) merger orbits on the other hand, tend to scatter stars in radial orbits resulting in more oblate and triaxial shapes as can be seen in Fig. \ref{fig:triax} (yellow markers). We conclude that merger remnants have less spherical shapes than isolated clusters, the process driven by the eccentricity of the progenitor cluster orbits.

\subsection{Star cluster kinematics}
\label{sec:kin}
In a spherical coordinate system, the two angular velocity dispersion components $\sigma_\mathrm{\theta}, \sigma_\mathrm{\phi}$ can be combined into a tangential velocity dispersion $\sigma_\mathrm{t}$ as

\begin{equation}
\sigma_{\mathrm{t}}=\sqrt{\frac{\sigma_\theta^2+\sigma_\phi^2}{2}}.
\end{equation}

The non-isotropic nature of the kinematic structure of the stellar system can be characterized by the velocity anisotropy parameter $\beta$ \citep{Binney2008}, defined as 

\begin{equation}
\beta=1-\frac{\sigma_\theta^2+\sigma_\phi^2}{2 \sigma_{\mathrm{r}}^2}=1-\frac{\sigma_{\mathrm{t}}^2}{\sigma_{\mathrm{r}}^2}.
\end{equation}

The velocity anisotropy $\beta$ is closely related to the orbital structure of a stellar population. When the stellar orbits are purely radial, then $\sigma_\mathrm{t}=0$ and $\beta=1$. Conversely, for a stellar population with exclusively circular orbits, $\sigma_\mathrm{r}=0$ and $\beta= -\infty$. The system is isotropic ($\beta=0$) if the two components $\sigma_\mathrm{r} , \sigma_\mathrm{t},$ are equal. 

\begin{figure*}
	\includegraphics[width=0.8\textwidth]{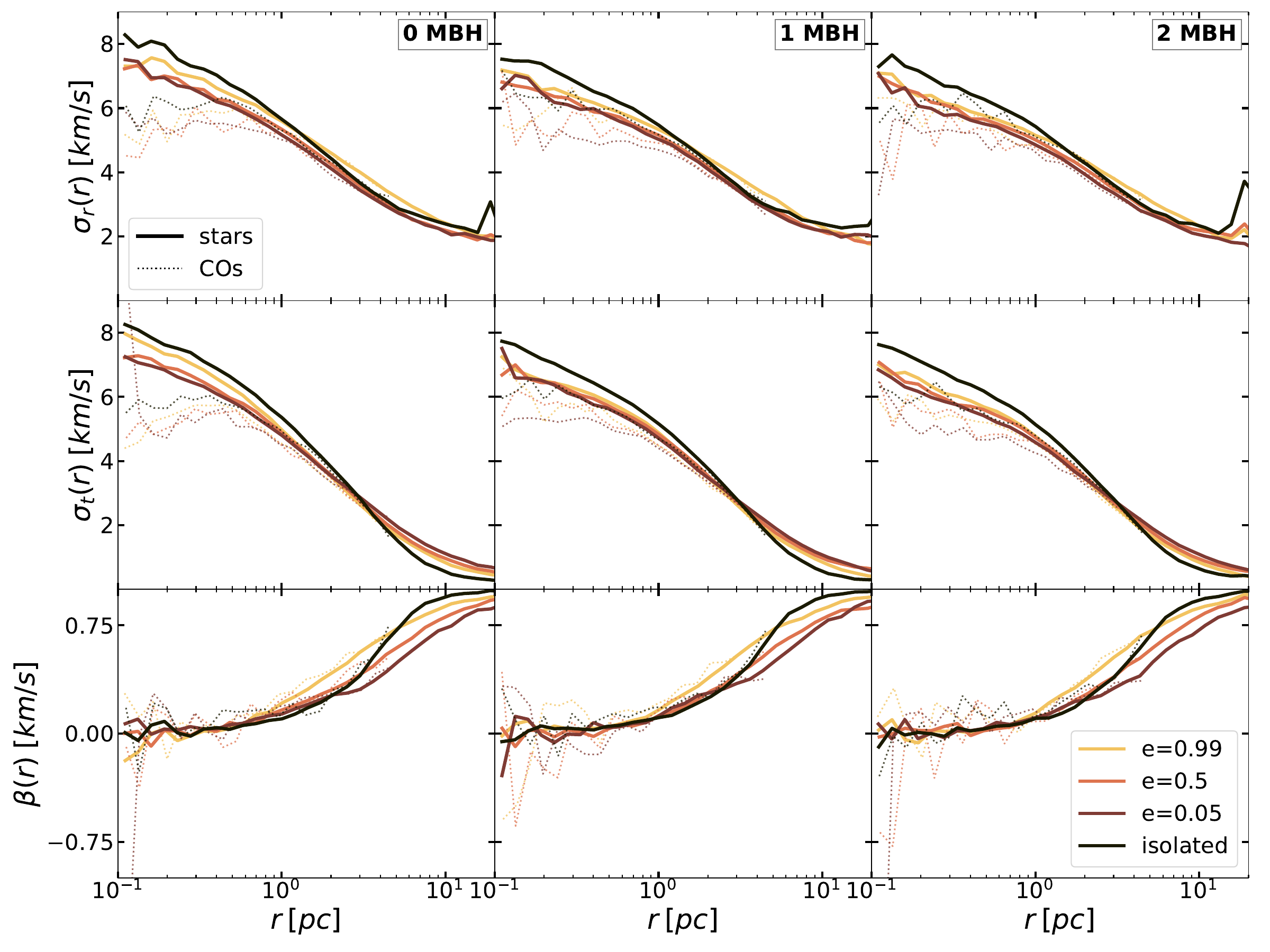}
    \caption{Radial $\sigma_\mathrm{r}$ (top panels) and tangential $\sigma_\mathrm{\phi}$ (middle panel) velocity dispersion profiles and the anisotropy parameter $\beta$ (bottom panels) for simulations without (left column) with one (middle column) and with two (right column) MBHs after 100 Myr. The thick solid lines correspond to stars only, while the thin dotted lines to the contribution of compact objects. All systems show isotropic velocity dispersions in the centre and more massive particles, i.e., compact objects have lower velocity dispersion, as expected from energy equipartition.}
    \label{fig:dispersion_final}
\end{figure*}

To compute velocity dispersion profiles for our star cluster setups and to reduce the noise we construct average dispersion profiles. To do so we use 10 simulation snapshots from the last $5\times 10^4 \: \mathrm{yr}$ in our runs. Each snapshot is centred in space and velocity to the region of highest stellar density using the shrinking sphere approach \citep{2003MNRAS.338...14P}. The velocity dispersion and anisotropy profiles (Fig. \ref{fig:dispersion_final}) of stars (thick solid lines) and of COs (thin dotted lines) of the remnants at and $t=100 \; \mathrm{Myr}$ are presented in Fig. \ref{fig:dispersion_final}. The left panels correspond to simulations without MBHs. Both $\sigma_\mathrm{r}$ and $\sigma_\mathrm{t}$ have values of about $\lesssim 10 \mathrm{~km} \mathrm{~s}^{-1}$ and decrease with increasing radius. The $\beta \sim 0$ profiles indicate isotropic velocity dispersion in the inner parts of all clusters. The merger orbits do not change the dispersion profiles significantly. The presence of one (middle column) or two (right column) MBHs also does not change the kinematic structure significantly. 

\subsubsection{Rotation of cluster merger remnants}
\label{sec:rotation}

For the simulated cluster mergers, orbital angular momentum is transferred to internal angular momentum. The effect is strongest for low eccentricity orbit (i.e. $e=0.05,0.5$) which have the highest orbital angular momentum. This corresponds to higher merger remnant rotation velocities.
\begin{figure*}
	\includegraphics[width=0.8\textwidth]{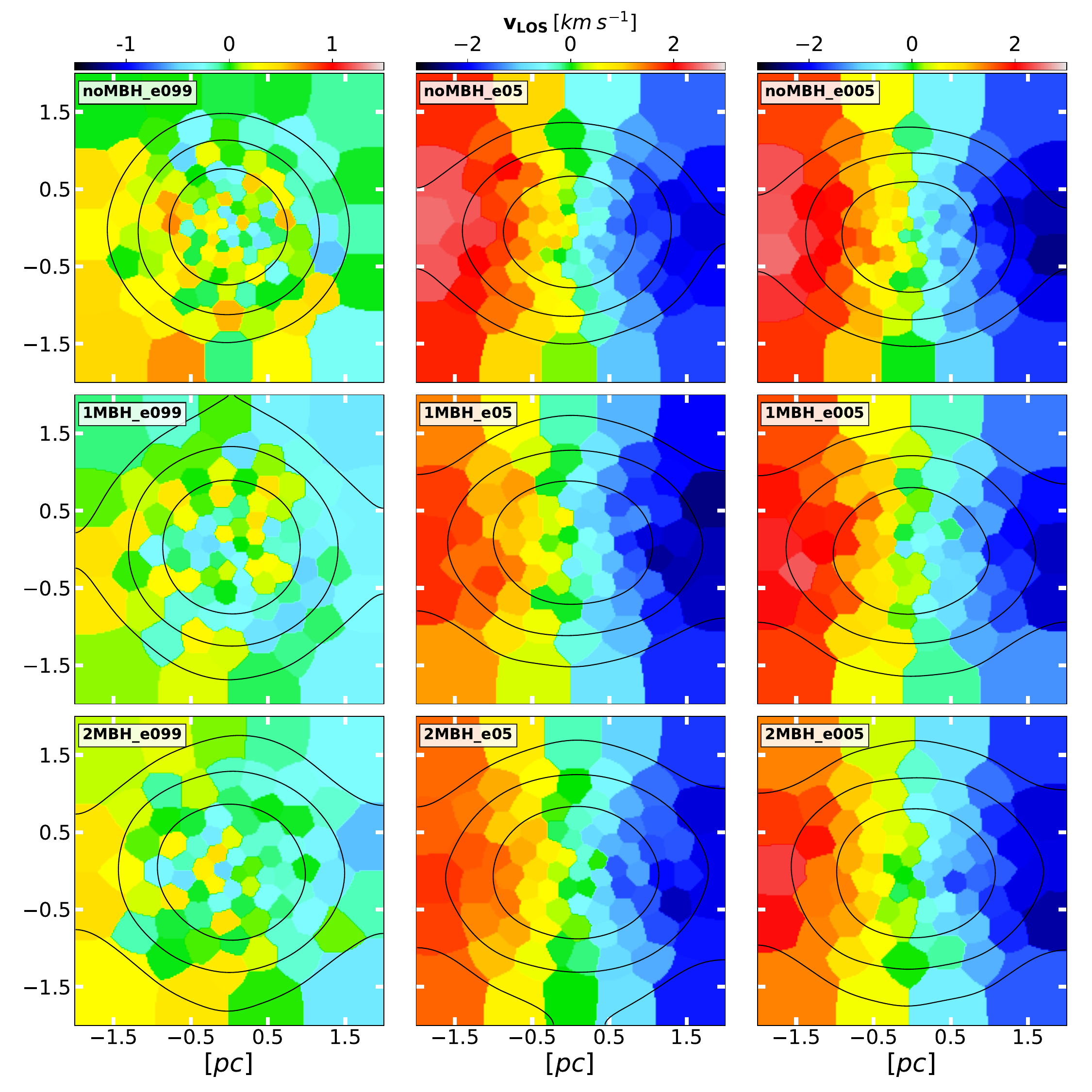}
    \caption{Two dimensional (x vs. z) line-of-sight velocity maps of the stars in the simulated cluster mergers at $t=100 \:\mathrm{Myr}$. From top to bottom we show the models with no, one, and two MBHs and the eccentricity of the merger orbit is decreasing from left (most radial) to right (most circular). The more circular merger remnants (middle and left panels) show no clear signs of rotation while the radial orbit remnant (left panels) show no rotation. The presence of an MBH does not affect the rotation properties. Contours of constant surface density are also displayed.}
    \label{fig:vlos_map}
\end{figure*}

We present the two dimensional line-of-sight (LOS) velocity ($V_{\mathrm{LOS}}$) maps of the cluster merger simulations after 100 Myr of evolution in Fig.  \ref{fig:vlos_map}. The maps are constructed utilizing the analysis package \pygad{}  \citep{Roettgers_2020}.  Following \citet{Naab_2014}, these maps are produced similar to the analysis of observational integral field unit data \citep[see e.g.][for more details]{Naab_2014}. We use the clusters' reduced moment of inertia tensor given by Eq. \ref{eq:inertia_tensor} to align the major axis of the cluster with the $x-$axis and perform the analysis within the central region of about $\sim 2$ half-mass radii, corresponding to a spatial extent of $4 \: \mathrm{pc}$. The velocity data is represented by \textit{spaxels} of constant signal-to-noise ratio (a fixed number of stars per spaxel) using a Voronoi tessellation algorithm \citep{Cappellari_2003}. We also show contours of constant surface density.

The highly eccentric ($e$ = 0.99) merger orbits carry the least angular momentum and the respective remnants show no sign of ordered rotation (left panels in Fig. \ref{fig:vlos_map}) independent of whether the clusters host black holes or not. The mergers with higher angular momentum orbits produce remnants which clearly show ordered rotation with velocities up to $\sim$ 3 $\mathrm{km} \: \mathrm{s}^{-1}$ (middle and right panels of Fig. \ref{fig:vlos_map}). In contrast to the velocity dispersion (see Sec. \ref{sec:kin}), the presence of MBHs in the clusters does not change the rotation features.

\section{Formation and Evolution of Massive Black Hole Binaries}
\label{sec:MBHB_evol}

In this section we focus on the formation and evolution of the MBH binaries in the merger runs \twoc{}, \twob{} and \twoa{} with two MBHs and the isolated cluster, \isotwo{}, with a central MBH binary. The different cluster orbits affect the time when the MBH binaries become bound in the cluster mergers, while the \isotwo{} MBH binary is bound by definition. For the \twob{} and \twoc{}  runs, the MBH binaries become bound at $1 \: \mathrm{Myr}$ and $1.3 \: \mathrm{Myr}$, respectively. For the merger with the highest eccentricity of $e$ = 0.99 (\twoa) the MBH binary becomes first bound at $1.1 \: \mathrm{Myr}$  then unbound again at $1.5 \: \mathrm{Myr}$. During the first bound phase the semi-major axis of the binary shrinks from $a_{\mathrm{b}} \approx 1 \: \mathrm{pc}$ down to $0.1 \: \mathrm{pc}$. The binary reaches a steady bound state at $\sim 3 \: \mathrm{Myr}$ with $a_{\mathrm{b}} \leq 0.1 \: \mathrm{pc}$. Interactions with the stellar background lead to a continuous energy exchange with the MBH binary. This leads to orbital energy being removed from the binary, resulting in decrease of their semi-major axes $a_{\mathrm{b}}$. 
When the MBH binary separation becomes lower than a critical value $a_\mathrm{h}$ of the semi-major axis, the binary has reached the so called 'hard' phase at \cite{Quinlan_1996}
\begin{equation}
    a_h \equiv \frac{G M_{\mathrm{2}}}{4 \sigma^2},
\end{equation}
where $M_{\mathrm{2}}$ is the mass of the secondary MBH (in our case $M_{1}=M_{2}=500 \: \mathrm{M_{\odot}}$) and $\sigma$ is the local velocity dispersion inside the $r_{\mathrm{SOI}}$ of the MBH binary.
A typical value for the hard separation in our simulations is $a_\mathrm{h} \approx 2\times 10^{-3} \: \mathrm{pc}$. The first time the MBH binaries in our simulations pass this threshold is at $t_{\mathrm{hard}}=1.7, \: 4.0,\: 4.8$ and $7.5 \; \mathrm{Myr}$  for the \isozero, \twoc, \twob, and \twoa{} models, respectively. We note that on galactic scales, the eccentricity $e_{\mathrm{b}}$ of formed bound SMBH binaries is strongly affected by stochastic effects \citep{Rawlings_2023}.

\begin{figure}
  \centering
  \includegraphics[width=0.9\columnwidth]{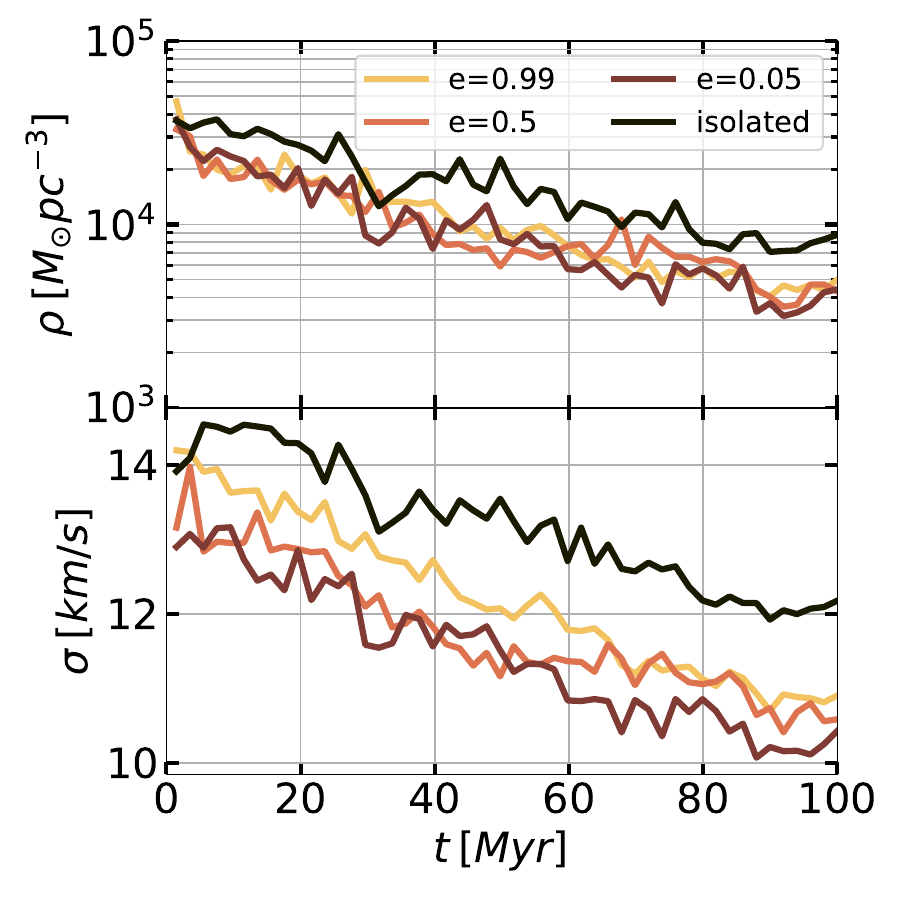}
  \caption{Time evolution of the density (top panel) and velocity dispersion (bottom panel) for simulations with MBH binaries. The values are computed within the sphere of influence $r_{\mathrm{SOI}}$ of the MBHs.}
  \label{fig:MBHB_sigrho}
\end{figure}

\subsection{Hardening and eccentricity growth of MBH binaries in a stellar background}

\begin{figure}
  \centering
  \includegraphics[width=0.9\columnwidth]{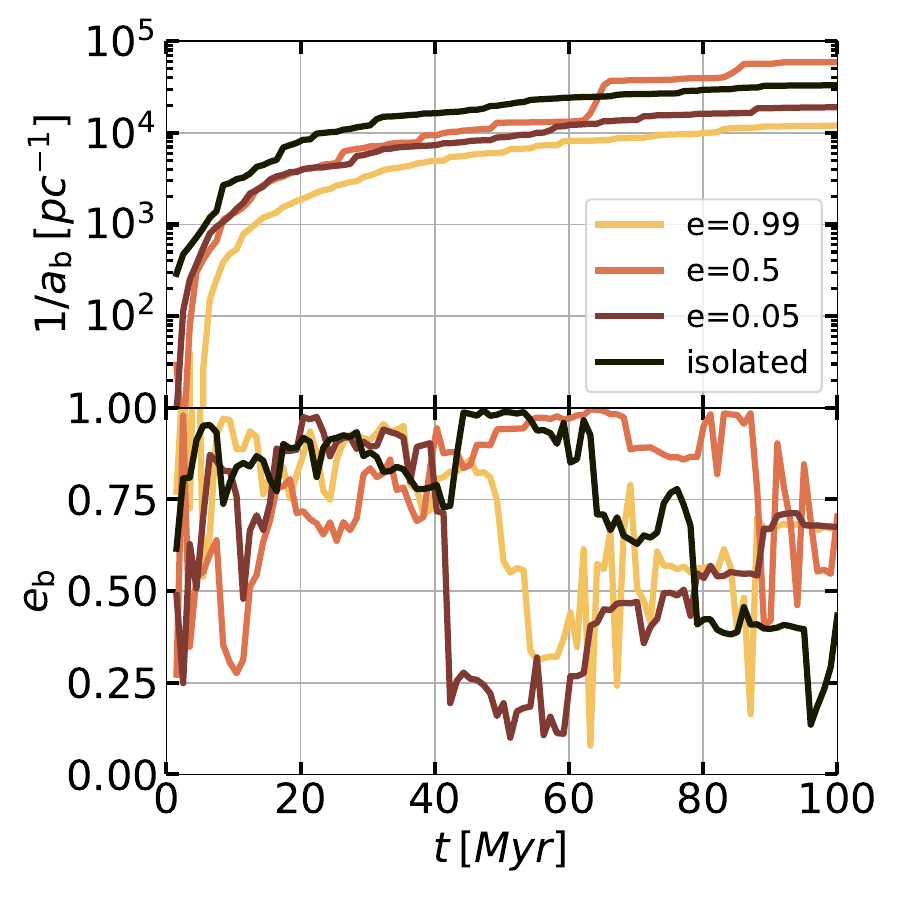}
  \caption{Time evolution of the MBH binaries' inverse semi-major axis $1/a_{\mathrm{b}}$ (top panel) and eccentricity $e_{\mathrm{b}}$ (bottom panel) evolution of the MBH binaries. The semi-major axis is monotonously decreasing due to continuous interactions with the central MBH binaries while the eccentricity shows strong variations.}
  \label{fig:MBHB_evol}
\end{figure}

The hardening of an MBH binary depends on the properties of its stellar environment, i.e., the velocity dispersion and mass density of the stars. For a fixed stellar background, the hardening rate $H$ and eccentricity growth $K$ are given by \citep{Quinlan_1996}, 
\begin{equation}
\begin{aligned}
H & =\frac{\sigma}{G \rho} \frac{\mathrm{~d}}{\mathrm{~d} t}\left(\frac{1}{a_{\mathrm{b}}}\right), \\
K & =\frac{\mathrm{d} e_{\mathrm{b}}}{\mathrm{~d} \ln (1 / a_{\mathrm{b}})},
\end{aligned}
\label{eq:hardening_rate}
\end{equation}
where $\rho$ and $\sigma$ are the stellar density and velocity dispersion inside the sphere of influence $r_{\mathrm{SOI}}$ (Eq. \ref{eq:rSOI}) of the binary. Equation \ref{eq:hardening_rate} assumes fixed values for $\rho$ and $\sigma$ and has been used in various studies on the efficiency of MBH hardening and its effect on the production of gravitational wave merger events and the production of RAs and HVs \citep{Sesana_2006,Sesana_2008,Sesana_2009,Leigh_2017,Rasskazov_2019, Gualandris_2022,Evans_2023}. In our simulations, ongoing dynamical interactions of the MBH binaries with the stellar background change the average central stellar density and velocity dispersion continuously. In Fig. \ref{fig:MBHB_sigrho} we show that the density inside the sphere of influence (top panel) of the MBH binaries continuously decreases with time, by a factor 4 - 5 during the 100 Myr of evolution. Over the same time interval, the stellar velocity dispersion (bottom panel in Fig. \ref{fig:MBHB_sigrho}) is decreased by $4 \: \mathrm{km}\:\mathrm{s}^{-1}$ for the merger remnants. The decrease in central density is caused by the ejection of stars which have interacted with the central MBH binary whose semi-major axis is continuously shrinking for all simulations with MBH binaries (the top panel of Fig. \ref{fig:MBHB_evol}). Individual strong encounters with the central MBH binary result in visible jumps (see e.g. the e = 0.5 merger, orange line, in the top panel of Fig. \ref{fig:MBHB_evol}) in the semi-major axis evolution and result in the eject of high velocity objects (see discussion in Sec. \ref{sec:ejection}).

The evolution of the eccentricity of the central MBH binary is not smooth (bottom panel of Fig. \ref{fig:MBHB_evol}) and fluctuates significantly. This adds uncertainties on the further post-Newtonian evolution of the MBH binary as the eccentricity has a strong impact the process of coalescence through gravitational wave emission \citep[e.g.][]{Bonetti_2020,Gualandris_2022}{}{} and may also imprint distinct signatures on the number and velocity distribution of captured or ejected stars \citep{Sesana_2006,Sesana_2008,Rasskazov_2019}.
We estimate the hardening rate coefficients $H$ and $K$ from the slope of linear fits on the $1/a_{\mathrm{b}}$ and $e_{\mathrm{b}}/\mathrm{ln}(1/a_{\mathrm{b}})$ time evolution. The values are given in Tab. \ref{tab:MBHB_properties}. The coefficients $H$ and $K$ have been typically evaluated (e.g. \citealt{Sesana_2006}) using scattering experiments assuming a fixed stellar background.

\begin{table}
\centering
\renewcommand{\arraystretch}{1.1} 
\setlength{\tabcolsep}{4pt} 
\small 
\begin{tabularx}{\columnwidth}{l|>{\centering\arraybackslash}X>{\centering\arraybackslash}X>{\centering\arraybackslash}X>{\centering\arraybackslash}X>{\centering\arraybackslash}X>{\centering\arraybackslash}X}
\hline
\hline
simulation & $H$ & $K$ & $a_{\mathrm{100}}$ & $e_{\mathrm{100}}$ & $\rho_{\mathrm{100}}$ & $\sigma_{\mathrm{100}}$ \\
           &     &  [$10^{-4}$]  & [$10^{-5}  \mathrm{pc}$] & & [$ M_{\odot}\mathrm{pc}^{-3}$] & [km/s] \\
\hline
2MBH\_e099 & 134.5 & -11  & 8.3 & 0.667 & 5000  & 12.15 \\
2MBH\_e05  & 263.2 & -9.6 & 1.7 & 0.704 & 4300 & 20.7  \\
2MBH\_e005 & 204.7 & -7.1 & 5.3 & 0.675 & 4400 & 10.86 \\
2MBH\_iso  & 352.3 & -8.7 & 3.0 & 0.433 & 8800 & 16.53 \\
\hline
\end{tabularx}
\caption{Hardening rate $H$ and eccentricity growth $K$ from linear fitting. $a_{\mathrm{100}}$, $e_{\mathrm{100}}$ are the semi-major axis and eccentricity of MBH binaries. $\rho_{\mathrm{100}}$, $\sigma_{\mathrm{100}}$ are density and velocity dispersion inside $r_{\mathrm{SOI}}$ at $t=100$ Myr.}
\label{tab:MBHB_properties}
\end{table}

Eccentricity growth has been observed again in previous studies \citep{Amaro_Seoane_2006, Amaro_Seoane_2009,Arca_Sedda_2019} in the context of merging star and globular clusters hosting IMBHs. It's origin can be due to the collective effect of many stars interacting with the MBHB or through strong encounters with individual and/or binary stars \citep{Askar_2021}.

\begin{figure}
  \centering
  \includegraphics[width=0.9\columnwidth]{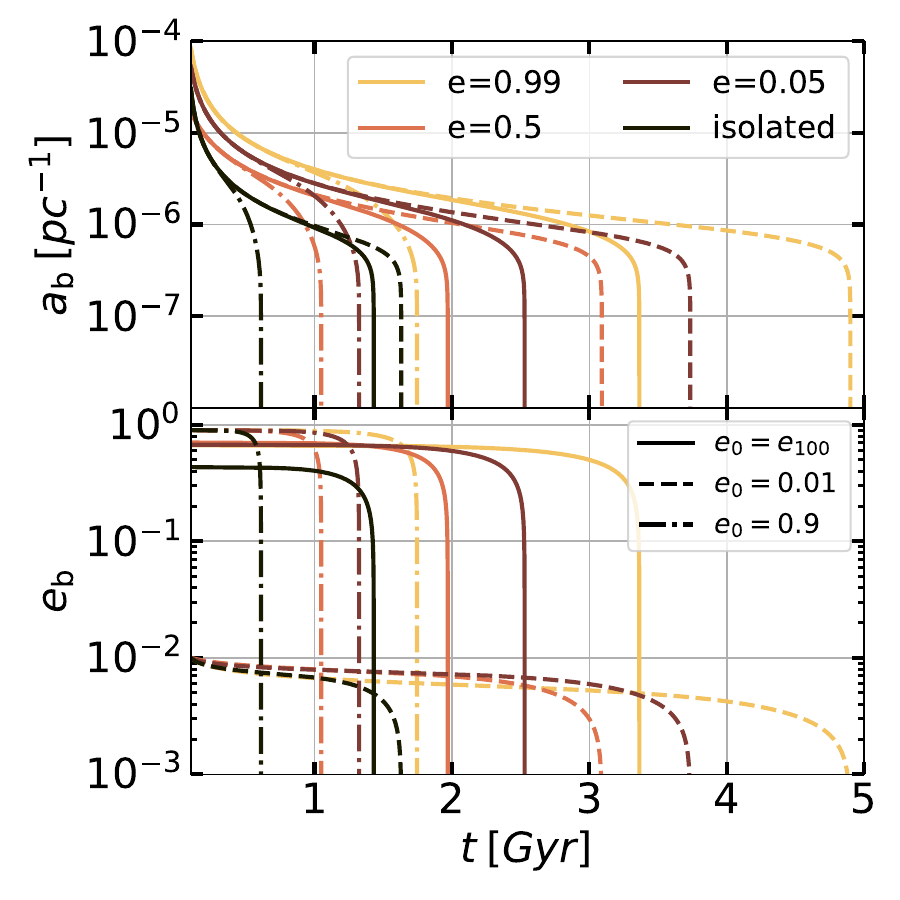}
  \caption{Predicted time evolution of the MBH binaries' orbital elements beyond $t=100 \: \mathrm{Myr}$ until coalescence. The solid lines corresponds to final values of MBHB eccentricities, therefore $e_{\mathrm{100}}$. We repeat the integration of Eq. \ref{eq:gw_odes} for nearly circular orbits with $e=0.01$ (dashed lines) and very eccentric ones $e=0.01$ (dot-dashed lines) , to test whether the binaries would still merge. All MBH binaries merge within $t<6 \: \mathrm{Gyr}$, with larger values of $e$ leading to earlier coalescence times.} 
  \label{fig:ae_gw}
\end{figure}

\subsection{Gravitational wave driven evolution and coalescence of MBH binaries}
In this section we explore the evolution of our MBH binaries after $t=100 \: \mathrm{Myr}$ where our simulations are stopped. To do so, we assume that the orbital elements evolve due to interactions of the MBH binary with the stellar background and GW emission which dominates at late times. The differential equations describing the evolution of semi-major axis and eccentricity are therefore written as 

\begin{equation}
\begin{aligned}
    & \frac{d a_{\mathrm{b}}}{d t}=\left.\frac{d a_{\mathrm{b}}}{d t}\right|_{\star}+\left.\frac{d a_{\mathrm{b}}}{d t}\right|_{\mathrm{GW}}, \\
    & \frac{d e_{\mathrm{b}}}{d t}=\left.\frac{d e_{\mathrm{b}}}{d t}\right|_{\star}+\left.\frac{d e_{\mathrm{b}}}{d t}\right|_{\mathrm{GW}},
\end{aligned}
\label{eq:gw_odes}
\end{equation}

where the terms with $\star$ refer to stellar interactions and $GW$ to GW-emission respectively. The former are described by \citet{Quinlan_1996},

\begin{equation}
\begin{aligned}
    \left.\frac{d a_{\mathrm{b}}}{d t}\right|_{\star} & =-a_{\mathrm{b}}^2 \frac{H G \rho}{\sigma} \\
    \left.\frac{d e_{\mathrm{b}}}{d t}\right|_{\star} & =a_{\mathrm{b}} \frac{H K G \rho}{\sigma},
\end{aligned}
\end{equation}

where $G$ is the gravitational constant and $\rho$ and $\sigma$ are the stellar density and velocity dispersion inside the sphere of influence $r_{\mathrm{SOI}}$ (Eq. \ref{eq:rSOI}) of the binary as before and $H$, $K$ correspond to the fitted values for the hardening rate and eccentricity growth given in Table \ref{tab:MBHB_properties}. The GW-driven part is modelled via Eq. \ref{eq: secular_a_ecc} where, $a,e$ correspond to the binary elements $a_{\mathrm{b}},e_{\mathrm{b}}$.

The initial values used for solving Eq. \ref{eq:gw_odes} correspond to the values at the end of the simulation, i.e. $(a_{\mathrm{100}},e_{\mathrm{100}}, \rho_{\mathrm{100}},\sigma_{\mathrm{100}} )$. The results are presented in Fig. \ref{fig:ae_gw} (solid lines), where all MBH binaries merge before $t<4 \: \mathrm{Gyr}$. Although $e_{\mathrm{100}}$ has some fixed value for each simulation but due to the stochastic nature of strong encounters, this value would vary if the simulation runs were slightly longer. For this reason and in order to capture the effect of $e_{\mathrm{b}}$ in the full range of possible values, we integrate system eq. \ref{eq:gw_odes} for an almost circular and a highly eccentric binary, whereas the $e_{\mathrm{100}}$ from the actual runs serve as intermediate ones. Assuming a fixed high initial eccentricity $(e_{\mathrm{b}}=0.9)$ the binaries merge earlier, in less than 2 Gyr. For almost circular orbits $(e_{\mathrm{b}}=0.01)$ the MBH binaries can take up to $\sim$5 Gyr to merge. The above estimates indicate that the MBHs formed in mergers of dense stars clusters will merge within a Hubble time. The main takeaway from this experiment is that given a sufficiently dense environment, hard MBH binaries could potentially merge regardless of how eccentric they are and can be observed with future space-born and ground-based GW detectors like LISA \citep{Amaro_Seoane_2006,Amaro_Seoane_2017_LISA,Arca_Sedda_2019} and Einstein Telescope \citep{Maggiore_2020}.

\section{The ejection of stars and compact objects}
\label{sec:ejection}
\subsection{Classification and overview}
In each simulation, we monitor the stars and compact objects (COs) which become unbound and eventually escape the cluster at a fixed distance of $r_\mathrm{esc}=100 \: \mathrm{pc}$. As outlined in Sec. \ref{section: ejections}, the various dynamical ejection channels have characteristic ejection velocities. In this section, we give an overview of the properties of ejected stars and COs in our simulations and connect them to the most likely ejection mechanisms. We classify the ejected objects as follows:

\begin{itemize}
  \item Low-velocity escapers ($v_{\mathrm{ej}} < 50 \mathrm{~km} \mathrm{~s}^{-1}$), originating mostly from weak 2-body encounters and 2-body relaxation in the vicinity of an MBH.
  \item Runaway stars (RA) and COs ($ 50 \mathrm{~km} \mathrm{~s}^{-1} \leq v_{\mathrm{ej}} < 300 \mathrm{~km} \mathrm{~s}^{-1}$), most likely from from 3- or 4-body encounters of single or binary stars with  BH+star and BH+BH or MBH+star and MBH+MBH binaries and/or binary encounters with a single MBH (a Hills-like mechanism) or a binary MBH.
  \item Hyper-runaway (HR) stars and COs ($ 300 \mathrm{~km} \mathrm{~s}^{-1} \leq v_{\mathrm{ej}} < 700 \mathrm{~km} \mathrm{~s}^{-1}$) , which are the extreme version of the previous channel from interactions with hard binaries. 
  \item Hyper-velocity stars (HV) and COs ($ v_{\mathrm{ej}} \geq 700 \mathrm{~km} \mathrm{~s}^{-1}$), most likely due to strong 3- and 4-body interactions with a single and/or binary MBHs.
  \item Extreme-velocity stars and COs ($ v_{\mathrm{ej}} \geq 1000 \mathrm{~km} \mathrm{~s}^{-1}$), which are a handful of extreme and rare cases for which we suggest the most likely scenario of ejection.
\end{itemize}

\begin{figure*}
    \centering
	\includegraphics[width=0.8\textwidth, keepaspectratio]{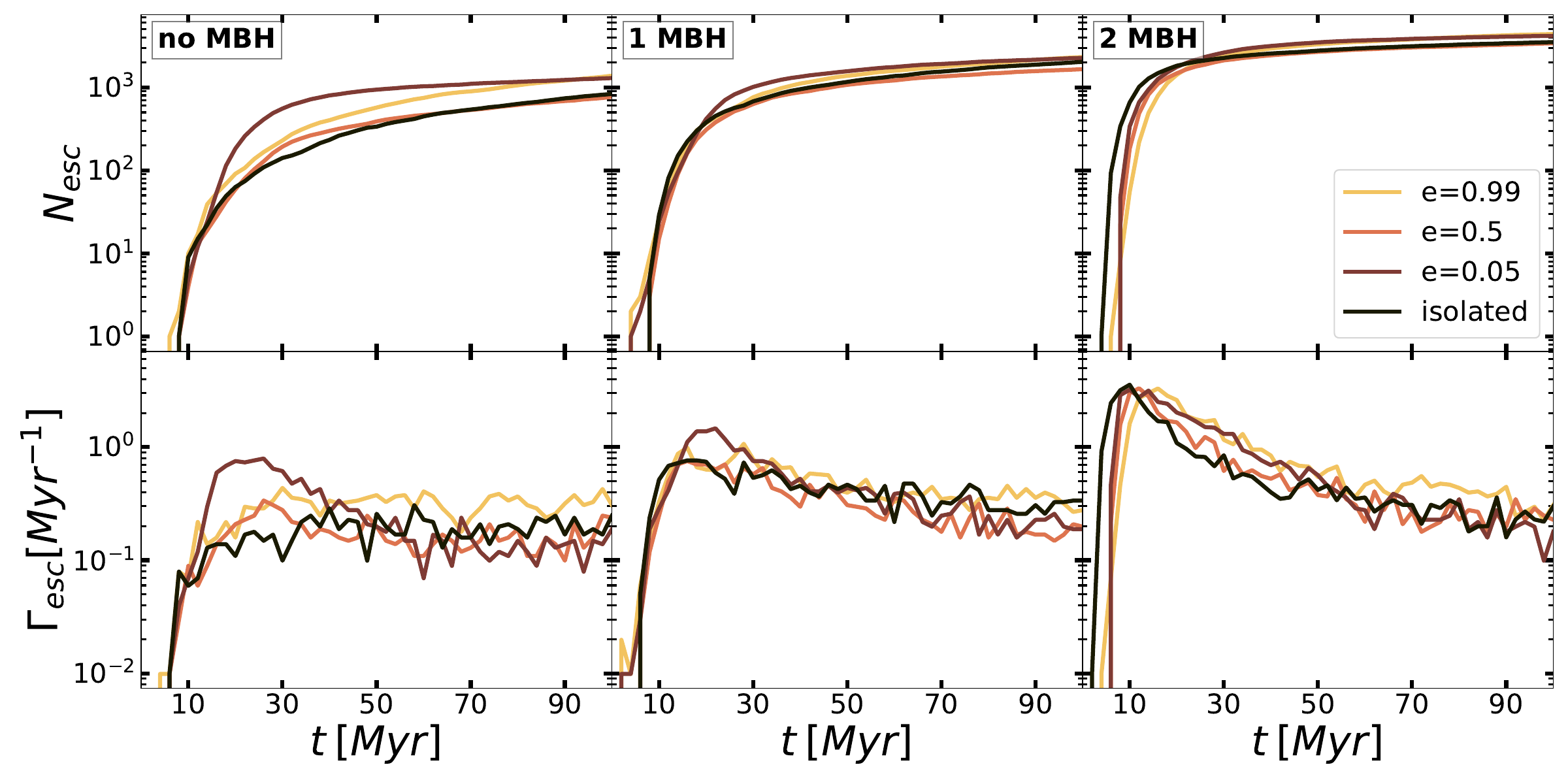}
    \caption{Time evolution of the cumulative number of escaping objects (stars and compact objects) at 100 pc (top panels) and the respective escape rate (bottom panels) of escaping stars and COs. For the binary MBH simulations the average (t < 30 Myr) escape velocity (up to 39 km s$^{-1}$) can be about a factor two higher than for the single MBH cases, which explains the earlier peak in the escape rates.}
    \label{fig:escapers}
\end{figure*}

\begin{figure*}
  \centering
  \includegraphics[width=0.8\textwidth]{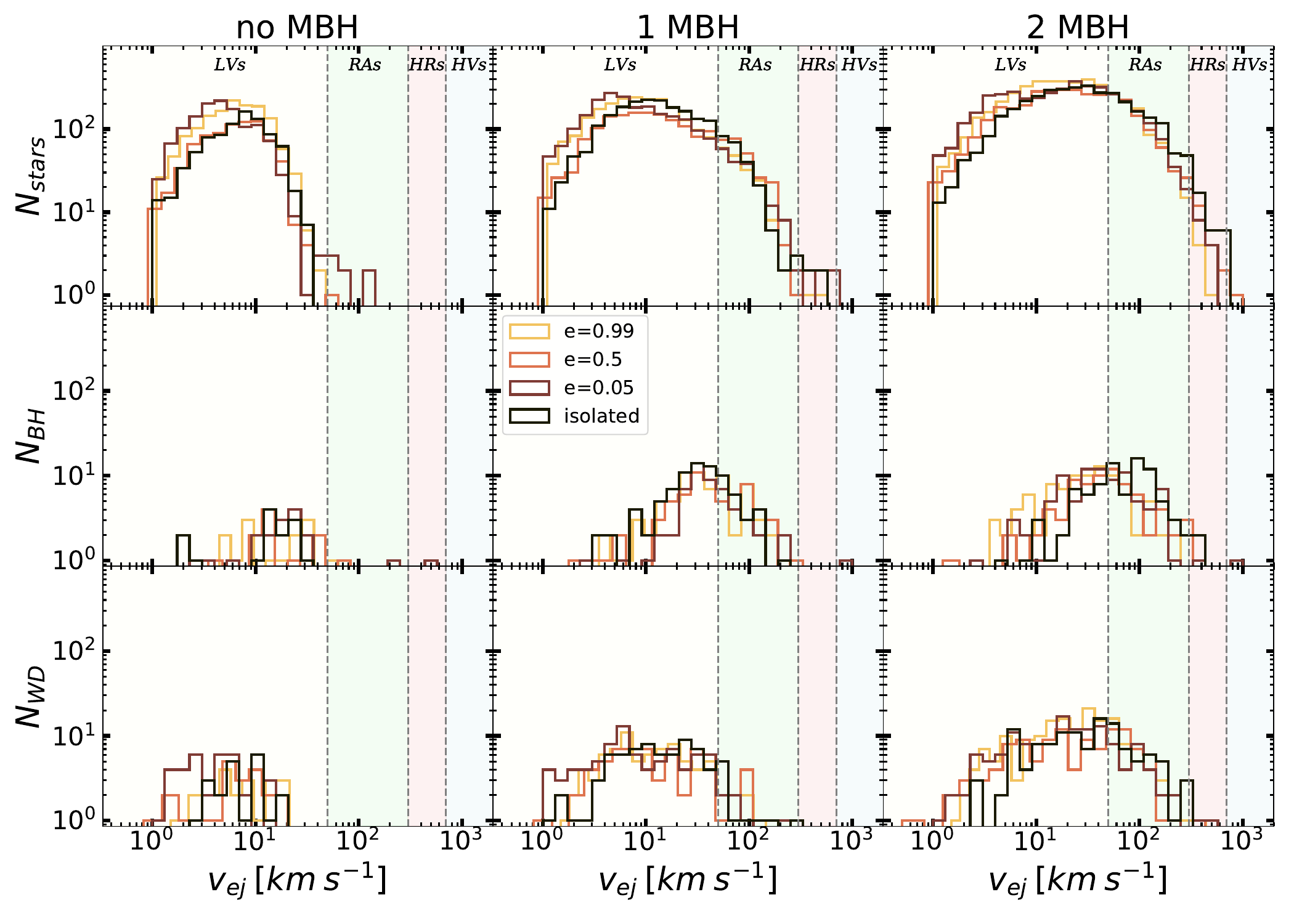}
    \caption{Distribution of ejection velocities for objects crossing 100pc distance for simulations without (left panels), with one 1000 $M_\odot$ (middle panels), and with two 500 $M_\odot$ MBHs (right panels). From top to bottom we show stars, stellar-mass BHs and WDs. The different colors indicate different orbital setups. The different shaded areas indicate the velocity regimes for runaway objects (RA,  $ 50 \mathrm{~km} \mathrm{~s}^{-1}\leq v_{\mathrm{ej}} < 300 \mathrm{~km} \mathrm{~s}^{-1}$), hyper-runaways (HR, $300 \mathrm{~km} \mathrm{~s}^{-1}\leq v_{\mathrm{ej}} < 700 \mathrm{~km} \mathrm{~s}^{-1}$), and hyper-velocity objects (HV, $ v_{\mathrm{ej}} \geq 700 \mathrm{~km} \mathrm{~s}^{-1}$), respectively. In particular the simulations with binary MBHs (right panels) produce a new population of stars, BHs, and WDs in the RA and even HV velocity regime.}
  \label{fig:eject-demogr}
\end{figure*}

\begin{figure*}
  \centering
  \includegraphics[width=0.95\textwidth]{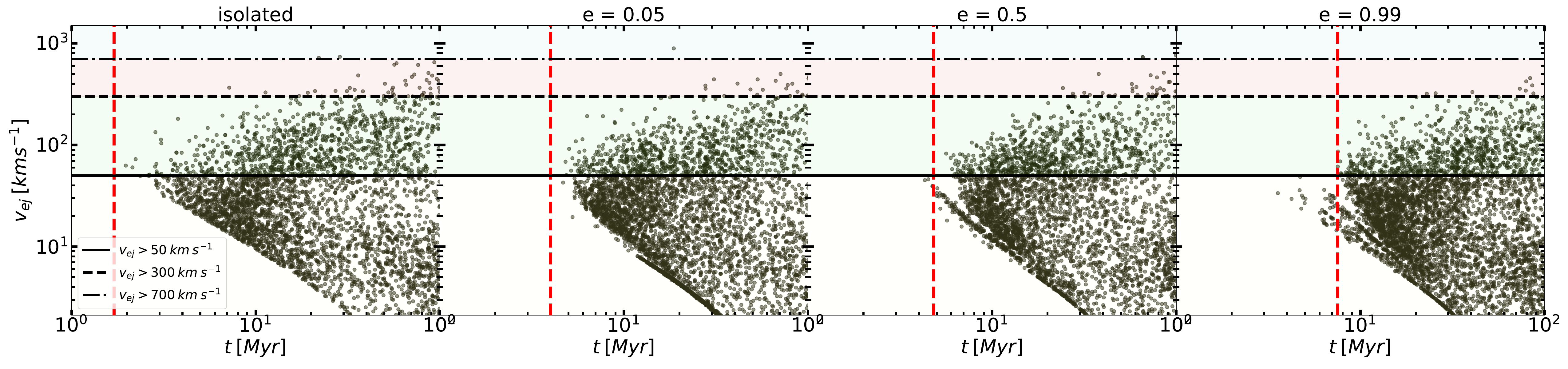}
    \caption{Distribution of ejection velocities versus time for stars and compact objects crossing a 100 pc sphere around the clusters with MBH binaries. The solid, dashed, and dashed-dotted horizontal lines indicate the velocity boundaries for low-velocity ejections ($v_{\mathrm{ej}} < 50 \mathrm{~km} \mathrm{~s}^{-1}$), runaway objects (RA,  $ 50 \mathrm{~km} \mathrm{~s}^{-1}\leq v_{\mathrm{ej}} < 300 \mathrm{~km} \mathrm{~s}^{-1}$), hyper-runaways (HR, $300 \mathrm{~km} \mathrm{~s}^{-1}\leq v_{\mathrm{ej}} < 700 \mathrm{~km} \mathrm{~s}^{-1}$), and hyper-velocity objects (HV, $ v_{\mathrm{ej}} \geq 700 \mathrm{~km} \mathrm{~s}^{-1}$), respectively. The red vertical lines indicate the time when the central MBH becomes hard. The lower velocity boundary is determined by the arrival time of the objects at a distance of 100 pc. The maximum escape velocities increase with time as the ejected objects result from interactions with more bound (more hardened) central MBH binaries (see Fig. \ref{fig:MBHB_evol}).}
  \label{fig:Vej_tot}
\end{figure*}

In Fig. \ref{fig:escapers} we show the evolution of the cumulative number of escapers (top panel) and the escape rate (bottom panel) for all simulations. The characteristic timescale for escaping our clusters can be estimated from the escaper removal distance and the cluster centre escape velocity yielding $100$ pc / $22 \mathrm{~km} \mathrm{~s}^{-1}$ $\sim4.4$ Myr. The total number of escapers raises rapidly after 10 - 15 Myr (the post merger phase) when the first stars cross the 100 pc boundary. After 100 Myr about 1000 stars have escaped for all simulations without MBHs. The escape rates (bottom left panel) are almost constant (though a moderate peak develops for the most circular merger) at $\sim 0.2/\mathrm{Myr}$. For simulations with one MBH (middle panels in Fig. \ref{fig:escapers}) the total number of escapers ($\sim$ 2000) is about a factor two higher compared to the run without MBHs, and the escape rates peak at  $\sim 1.0/\mathrm{Myr}$ after $\sim 20 \: \mathrm{Myr}$ . At the end of the simulation the escape rate drops to similar values as the no MBH simulations. In the presence of two MBHs (right panels) the total number of escapers raises to $\gtrsim$ 3000 objects and the escape rates peak earlier (due to the higher speeds of the escapers) at values of $\sim 3.0/\mathrm{Myr}$. After $100 \: \mathrm{Myr}$ the rates have dropped to similar values as the simulations with one and no MBH. Even though the escape rates are similar towards the end of the simulations, the velocities of the objects are significantly higher in the presence of MBHs. We will discuss this fact more below.  

The number $N{\mathrm{esc}}$ and rate $\Gamma_{\mathrm{esc}}$ of ejections do not strongly correlate with the orbital eccentricity of the mergers. The results indicate a weak dependence to our ICs, thus suggesting that the our escaper numbers and rate estimates are robust and a generic feature of star cluster mergers with and without MBHs. For clusters without MBHs, only the \zeroc{} remnant shows an increased number and rate of $1/\mathrm{Myr}$ of escapers in the first $15 \mathrm{Myr}$ compared to the ($\Gamma_{\mathrm{esc}}<0.5$ ejection per $\mathrm{Myr}$) other runs, which could be due to the higher number of dynamically formed binaries during that time (Section \ref{sec:binaries}). 
When MBHs are present, the merger orbit itself does not significantly affect the overall number and rates of ejection, as we see from the bottom panels (middle, right) in Fig. \ref{fig:escapers} where $\Gamma_{\mathrm{esc}}$ from the isolated runs is similar to those of the merger remnants. 

The velocity distribution of the escapers in all simulations are shown in Fig. \ref{fig:eject-demogr} separated into stars (top panels), stellar BHs (middle panel) and white dwarfs (WD) bottom panels. The assumed velocity limits for runaway stars, hyper-runaway stars and hyper-velocity stars are indicated. The total number of escaping stars, BHs, and WDs for these three classes are given in first four columns of Tab. \ref{tab:ejection_number} and the total mass and fraction of the respective population lost by escapers is given in the five columns on the right. For the isolated cluster without MBHs all escapers are low velocity escapers and we do not find stars with velocities in excess of the assumed runaway star limit of 50 km s$^{-1}$. Merger of clusters without MBHs can produce a small population of runaway stars ($\lesssim 10$) and, in rare cases, a hyper-runaway BH. Already in the presence of one MBH the isolated and merger simulations produce a sizable population of $\sim 250$ runaway objects, up to five hyper-runaway stars. The presence of a MBH also results in the ejection of a significant fraction of the initial BH population (from $\sim$ 22 to $\sim$ 37 per cent), some with runaway velocities, and a few percent of the stellar, white dwarf, and neutron star population in the runaway velocity regime. If both clusters have an MBH which is forming a binary in the remnants, the runaway population increases to $\sim 800$ and up to 45 hyper-runaway and 3 hyper-velocity stars can be formed. The number of the escaping BHs increases slightly. These results clearly indicate the existence of a single or binary MBH in a star cluster will create a new population of runway stars and compact objects. 

In Fig. \ref{fig:Vej_tot} we show the distribution of ejection velocities for objects crossing a distance of 100 $\mathrm{pc}$ as a function of time for all simulations with binary MBHs. The results are not sensitive to the choice of 100 $\mathrm{pc}$ as the particle removal radius, since this would only slightly affect (increased/decreased for lower/higher values) the low-end of the the $v_{\mathrm{ej}}$ distribution. After the central MBHs become a hard binary (horizontal red dashed lines) objects are ejected with velocities exceeding 50 km s$^{-1}$ which is the velocity limit for runway stars. The ejection velocity of a star ejected via a strong encounter with an MBH binary, typically scales with the orbital velocity of the binary $v_{\mathrm{b}}$ \citep{Valtonen_2006,Merritt_2013_book}

\begin{equation}
v_{\mathrm{ej}} \sim v_{\mathrm{b}}=\left(\frac{2 G M_{\bullet}}{a_{\mathrm{b}}}\right)^{1 / 2},
\end{equation}

where $M_{\bullet}=M_{1}+M_{2}$. At later times, when the central MBH binary becomes harder with $a_{\mathrm{b}} \lesssim 10^{-3} \: \mathrm{pc}$ (see Fig. \ref{fig:MBHB_evol}) the maximum escape velocities increase into the hyper-runaway velocity regime. The lower velocity boundary for data points in this figure is determined by the flight times to the 100 pc particle removal distance, with the particle ejection velocities from the central parts of the cluster.   

 \begin{table*}{
     \centering
     \begin{tabularx}{0.95\textwidth}{l|ccccccccc} 
         \hline
         \hline
         simulation & $N_{\mathrm{LV}}$ & $N_{\mathrm{RA}}$ & $N_{\mathrm{HR}}$ & $N_{\mathrm{HV}}$ &  $M_{\mathrm{tot,esc}}$ & $N_{\mathrm{BH,esc}}$ & $N_{\mathrm{NS,esc}}$ & $N_{\mathrm{WD,esc}}$ & $N_{\mathrm{stars}}$ \\
       & stars/BHs/WDs & stars/BHs/WDs & stars/BHs/WDs & stars/BHs/WDs & [\%]      & [\%]     & [\%]     & [\%]     & [\%]  \\
 \hline
           2MBH\textunderscore e099  & 3607 / 65 / 134 & 813 / 29 / 41  & 8 / 0 / 1  & 0 / 0 / 0 & 7.4 & 36.9 & 4.0 & 4.2 & 3.5 \\
           2MBH\textunderscore e05   & 2616 / 41 / 76  & 786 / 36 / 39  & 34 / 0 / 1 & 3 / 0 / 0 & 5.8 & 30.5 & 2.5 & 2.8 & 2.7 \\
           2MBH\textunderscore e005  & 3343 / 51 / 116 & 837 / 37 / 27  & 19 / 1 / 2 & 0 / 1 / 0 & 6.9 & 34.9 & 3.2 & 3.5 & 3.4 \\
           2MBH\textunderscore iso   & 2568 / 31 / 88  & 935 / 55 / 33  & 40 / 3 / 2 & 2 / 0 / 0 & 6.4 & 38.7 & 2.8 & 2.9 & 2.8  \\
           1MBH\textunderscore e099  & 2182 / 37 / 79  & 167 / 28 / 8   & 2 / 0 / 0  & 0 / 0 / 0 & 5.1 & 30.3 & 1.8 & 2.1 & 1.9 \\
           1MBH\textunderscore e05   & 1442 / 41 / 57  & 242 / 23 / 10  & 0 / 0 / 1  & 0 / 0 / 0 & 4.3 & 25.8 & 1.8 & 1.6 & 1.3\\
           1MBH\textunderscore e005  & 2118 / 37 / 79  & 167 / 18 / 6   & 5 / 0 / 0  & 0 / 1 / 0 & 4.2 & 22.7 & 1.4 & 2.0 & 1.8\\
           1MBH\textunderscore iso   & 1870 / 62 / 67  & 204 / 22 / 10  & 4 / 0 / 0  & 0 / 0 / 0 & 5.0 & 37.2 & 1.9 & 1.8 & 1.7 \\
           noMBH\textunderscore e099 & 1427 / 13 / 26  & 1 / 2 / 0      & 0 / 0 / 0  & 0 / 0 / 0 & 1.6 & 6.0 & 0.7 & 0.6 & 1.1 \\
           noMBH\textunderscore e05  & 805  / 10 / 22  & 1 / 1 / 0      & 0 / 0 / 0  & 0 / 0 / 0 & 1.1 & 4.4 & 0.6 & 0.5 & 0.6  \\
           noMBH\textunderscore e005 & 1308 / 16 / 36  & 7 / 1 / 0      & 0 / 1 / 0  & 0 / 0 / 0 & 1.7 & 7.2 & 0.5 & 0.8 & 1.1  \\
           noMBH\textunderscore iso  & 869  / 14 / 21  & 0 / 0 / 0      & 0 / 0 / 0  & 0 / 0 / 0 & 1.2 & 6.2 & 0.5 & 0.4 & 0.7       
    \end{tabularx}
}
 	\caption{Escaper demographics for all simulations. For each simulation  we give, from left to right, the number of stars, BHs, and WDs, in the LV, RA, HR, and HV velocity regime. The total escaped fraction in total mass and numbers of BHs, NSs, WDs, and stars are given in columns 6 to 10.}
     \label{tab:ejection_number}
\end{table*}

\begin{figure}
  \centering
  \includegraphics[width=0.8\columnwidth]{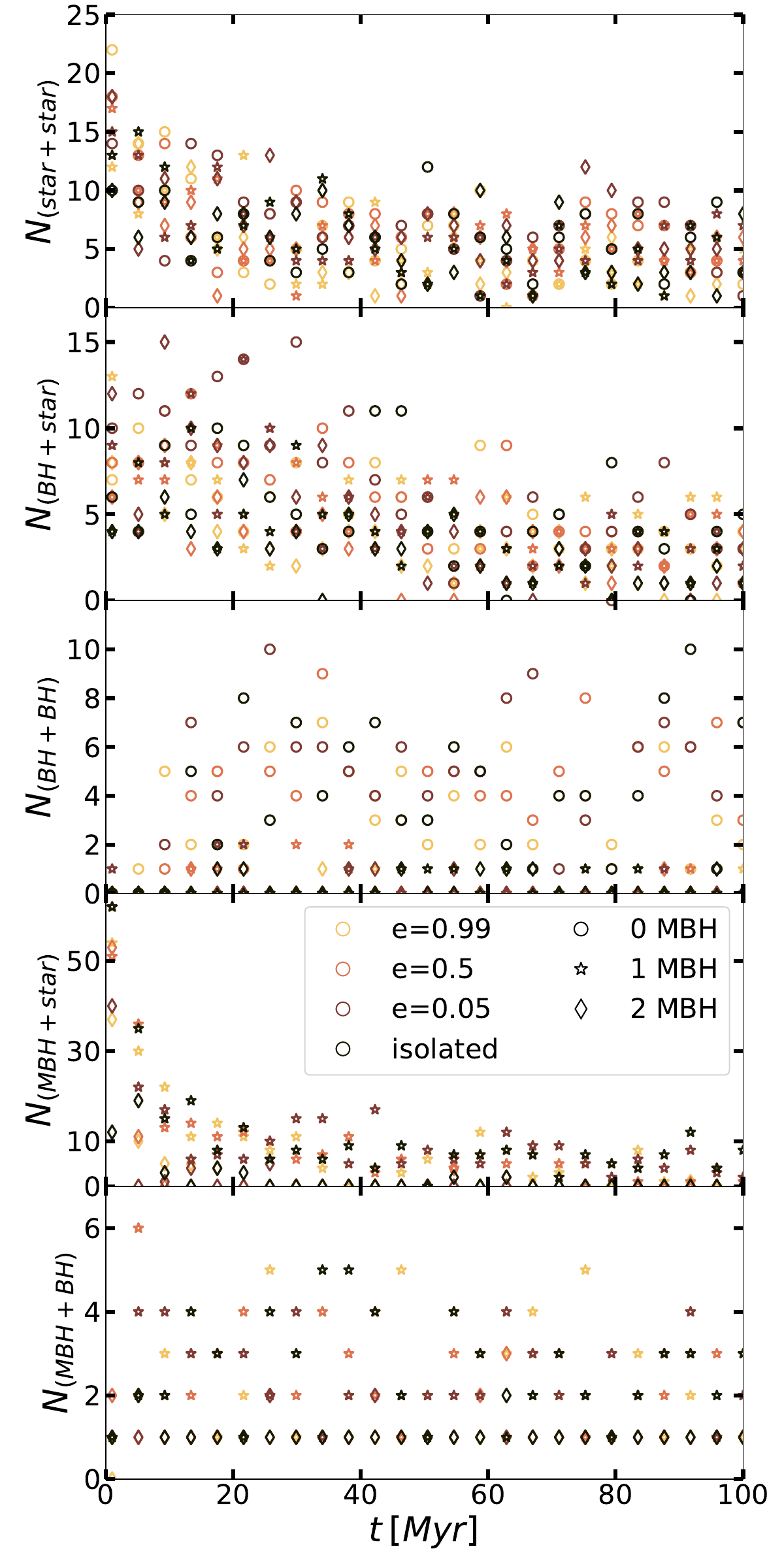}
  \caption{Number of dynamically formed binaries in all simulations as a function of time for snapshots with 4 Myr separation. From top to bottom we show star-star, stellar BH - star, stellar BH - BH, MBH - star, and MBH - BH (stellar) binaries. The simulations with MBHs hardly form any BH - BH binaries (third panel).}
  \label{fig:binaries}
\end{figure}

\subsection{Ejection mechanisms}

\subsubsection{Dynamically formed binary systems}
\label{sec:binaries}

Single-binary and binary-binary interactions play a central role in the ejections of stars from their host clusters. Besides from being formed in multiples, stars may form binary systems via dynamical processes, especially in dense cluster environments. These dynamical processes are present in our simulations and they include tidal captures (TC) \citep{Press_1977,Giersz_1986,Generozov_2018,Rizzuto_2023} or 3-body encounters known as \textit{three-body binary formation} (3BBF) \citep{Mansbach_1970, Aarseth_1976, Goodman_1993, Portegies_Zwart_2000a,Heggie_2003, Atallah_2024,Ginat_2024}. We briefly discuss the dynamical binary formation in Appendix \ref{apndx:A}.

We show the number of dynamically formed binaries for snapshots with 4 Myr separation in Fig. \ref{fig:binaries}. The initial conditions do not contain binary stars or compact objects. All simulations form stellar (low mass) binaries. Initially $(t<4\:\mathrm{Myr})$ up to 25 such systems form and the numbers drop to up to 10 later on throughout the runs (top panel) without a clear trend with initial conditions. For stellar BH+star binaries (second panel) the trend is similar with slightly lower (about 10 less on average) numbers. For stellar BH+BH binaries (third panel, Fig. \ref{fig:binaries}) there is a clear trend. Simulations with central MBHs have at most two such systems. Meanwhile, the simulations without central MBHs can have up to 10 BH-BH binaries. In these simulations, more BHs escape and binaries can be disrupted by the interaction with the central MBHs which leaves one BH directly bound to the central MBHs (see bottom panel). After $\sim$ Myr up to 10 stars are bound to the single MBHs and up to three to the binary MBHs (fourth panel). This trend also holds for MBH - BH binaries (bottom panel, Fig. \ref{fig:binaries}). Binaries with white dwarfs WD+WD or WD+star do form occasionally with an average number of $N_{\mathrm{WD+WD}} \approx N_{\mathrm{WD+star}} \approx 1-2$, but most of them do not survive due to multiple dynamical encounters and the fact that they form in relatively wide configurations.

\subsubsection{Ejections from clusters without MBHs}
First we examine the ejections from clusters without an MBH. In Fig. \ref{fig:escapers} we see these clusters mostly produce low-velocity escapers ($v_{\mathrm{ej}} \lesssim 10 \mathrm{~km} \mathrm{~s}^{-1}$) corresponding to 2-body relaxation and weak encounters between single stars. Still, there is a noteworthy population of ejections above $ 10 \mathrm{~km} \mathrm{~s}^{-1}$ which could be the result of encounters of single stars/COs with star+star, star+BH or BH+BH binaries (see Fig. \ref{fig:binaries}). An encounter with a star+star binary leads to velocity kicks of the same order as 2-body relaxation driven escapers, whereas encounters with star+BH or BH+BH binaries can explain the higher velocities. The range of velocities for these ejection processes strongly depends on the binary semi-major axis and mass ratio $q$. For example, for a hard binary with $a=2 \: \mathrm{AU}$ and $q=1/25$, Eq. \ref{eq:BHB} implies a velocity kick of $v_{\mathrm{ej}} > 100  \mathrm{~km} \mathrm{~s}^{-1}$. From Fig. \ref{fig:escapers} we notice that the \zeroc{} remnant is able to produce a few RAs even in the absence of an MBH as there is a sufficient number of BH+star and BH+BH binaries present in the simulations (Fig. \ref{fig:binaries}) to fuel the few-body ejection channels. Especially the hyper-runaway ($v_{\mathrm{ej}} \approx 473 \mathrm{~km} \mathrm{~s}^{-1}$) object, which is a BH (see middle-left panel in Fig. \ref{fig:escapers}), is ejected at $20\:\mathrm{Myr} < t < 25\: \mathrm{Myr}$. At this time the star cluster remnant contains $N_\mathrm{BH+star}=15$ and $N_\mathrm{BH-BH}=10$ which is the largest number of those types of binaries throughout the run. As such, the ejection time coincides with the most probable moment for a strong few-body ejection.

\subsubsection{Ejections from clusters with a single MBH}

Next we discuss ejections from clusters and remnants with a single MBH. In addition to the ejection channels in the case without any MBHs, now there is a significant fraction of MBH+star and MBH+BH binaries. The numbers of those binaries vary from $N_{\mathrm{MBH+star}}\approx 50$ at early times to $N_{\mathrm{MBH+star}} \gtrsim 10$ at later times and $N_{\mathrm{MBH+BH}}\approx 3-6$ (Fig. \ref{fig:binaries}) and essentially generate ejections covering the full range of $v_{\mathrm{ej}}$ (given a broad distribution of semi-major axes and mass spectrum of stars and BHs). The Hills mechanism also contributes and can explain some rare but strong encounters of binaries with the MBH. The number of stellar and BH binaries is significantly lower compared to the simulations without MBHs. The MBH+star and MBH+BH channel is likely dominant in this case. The number of Hills-like encounters is low due to the lack of available dynamically formed stellar-mass binaries. The number of RAs lies in the range of $192 < N_\mathrm{RAs} < 280$ for all cases, with a handful $(0-5)$ of HRs, and $0-2$ HVs. We see from Fig. \ref{fig:escapers} that none of them is a BH. Most HRs and stars with higher $v_{\mathrm{ej}}$ are produced in the \onec{} run, which has a larger number of MBH+star and MBH+BH binaries (Fig. \ref{fig:binaries}), implying stellar encounters with these types of binaries. 

A notable (classified as 'extreme' and not shown in Fig. \ref{fig:escapers}) ejection is one at $t \approx 87 \: \mathrm{Myr}$ of the \onec{} run with $v_{\mathrm{ej}} > 1000  \mathrm{~km} \mathrm{~s}^{-1}$. Specifically, a star of $m_\mathrm{\star}=0.7 \:\mathrm{M_{\odot}}$ accompanied by a $m_{\bullet}=14 \: \mathrm{M_{\odot}}$ BH, where the two used to be members of a bound binary system before the ejection. A MBH and star+BH interaction, i.e., a Hills ejection, is the most probable reason for this escape event. The binary semi-major axis in the last snapshot about $1000 \: \mathrm{yr}$ before the ejection was $a \approx 27 \: \mathrm{AU}$, which gives an estimated ejection velocity using Eq. \ref{eq:Hills} of $v_{\mathrm{ej}} \approx 900 \mathrm{~km} \mathrm{~s}^{-1}$. This ejection velocity, close to $1000 \mathrm{~km} \mathrm{~s}^{-1}$, is very well consistent with the Hills mechanism. We highlight the potential importance of the Hills mechanism for the production of high velocity stars in the context of star clusters and BHs in the IMBH mass regime as the Hills mechanism is commonly associated with galactic nuclei and their SMBHs.

\subsubsection{Ejections from clusters with a binary MBH}

Finally, we focus on ejections from clusters with a binary MBH at their centre. We note that although all the previous ejection channels are possible in those clusters, the vast majority does not involve interactions with stellar or BH binaries, since their numbers are very low (see Fig. \ref{fig:binaries}). The dominant additional ejection mechanism here is the interaction of an MBH+MBH binary with a star or a stellar BH. The MBH binaries become hard, i.e., $a_{\mathrm{b}} \leq a_{\mathrm{h}}$, already in the early $t < 5 \: \mathrm{Myr}$ post-merger phase as we discussed in Sec. \ref{sec:MBHB_evol}. The MBH binaries efficiently produce additional RAs, HRs and even HVs. Fig. \ref{fig:Vej_tot} marks (red dashed lines) the time at which the MBH binaries become hard, while the different horizontal (black) lines correspond to the various ejection velocities ($v_{\mathrm{ej}}$ is the velocity of the ejected object at $r_\mathrm{esc}=100 \; \mathrm{pc}$) classes. High-velocity ejections are initiated only after $a_{\mathrm{b}} = a_{\mathrm{h}}$. With enough available stars and COs in the vicinity of the MBH binary (top-right panel in Fig. \ref{fig:bound_rSOI}, the number of ejections rises from $1830 < N_{\mathrm{esc}} < 2450$  for the single MBH clusters to $3600 < N_{\mathrm{esc}} < 4800$  for the ones with an MBHB (top panels in Fig. \ref{fig:escapers}). 

\section{Summary and Conclusions}
\label{sec:Conclusions}
Star clusters in galaxies are expected to frequently interact and occasionally merge. This can happen during their hierarchical assembly or for evolved star clusters in the spiral arms of galactic discs, in galactic halos, or in galactic centres. In this study we explore the consequences of the merging star clusters hosting massive black holes for the cluster evolution and the production of escaping stars and COs. With $M_\bullet =500 \: \mathrm{M}_\odot$ the MBHs have masses about one order of magnitude higher than the most massive observed massive stellar BH \citep[see e.g.][]{2024A&A...686L...2G}. Such low mass MBHs might naturally form by rapid stellar collisions in hierarchically forming young and dense star clusters \citep[see e.g.][]{Gieles_2011,Rantala2024} and be retained in the clusters at later times. 

We have performed a suite of simulations of mergers of star clusters with individual cluster masses of $M_{\star} = 2.7 \times 10^4 \; \mathrm{M}_{\odot}$ and $N=64000$ individual stars and compact objects. In one set of simulations, each of the merging star clusters has a central MBH of $M_{\bullet} = 500 \: \mathrm{M}_\odot$. For comparison we perform simulations without initial MBHs, and with only one MBH of $M_{\bullet} =1000 \: \mathrm{M}_{\odot}$. We merge the clusters on a circular, a very radial and an intermediate eccentricity orbit. To estimate the effect of the merger process itself we also simulate isolated clusters with $M_{\star} = 5.4 \times 10^4 \: \mathrm{M}_{\odot}$ without any MBHs, with one MBH (1000 $\mathrm{M}_{\odot}$), and including a binary (2$\times 500 \: \mathrm{M}_\odot$) MBH. The simulations are performed with the hierarchical forward fourth-order, GPU accelerated N-body code \bifrost{} \citep{Rantala2023} which includes a regularisation scheme for binaries and post-Newtonian dynamics up to order PN3.5. We study the impact of the presence of MBHs on the star cluster merger remnants. Specifically we study how they affect the i) kinematic and structure of the remnant clusters, ii) the formation and evolution of MBH binaries and their coalescence time and iii) the production and populations of ejected stars and compact objects.

The clusters merge rapidly and the merger remnants are almost spherical with isotropic velocity dispersions. The two more circular merger orbits with $e=0.05$ and $e=0.5$ result in remnant star clusters with rotational velocities of $v_{\mathrm{LOS}}=v_{\mathrm{rot}}\sim 3 \mathrm{km}\:\mathrm{s^{-1}}$, similar to observations \citep[e.g.][]{Bellazzini_2012,Fabricius_2014}{}{} who find a few $\mathrm{km}\:\mathrm{s^{-1}}$. Apart from a slightly reduced central density of the order of $\sim 300-700 \: \mathrm{M_{\odot}\:\mathrm{pc^{-3}}}$, we find no strong evidence for a measurable/observable impact of the MBHs on the structural and kinematic properties of the merger remnants themselves. In conclusion, it's difficult to probe the presence of MBHs in merger remnants based on their kinematic properties alone.

In the cluster merger remnants the sinking MBHs rapidly form binaries and harden by interactions with stars and compact objects. This process produces a new population of escaping stars and compact objects with velocities $\gtrsim$ 50 km s$^{-1}$, which is absent in the star cluster mergers without MBHs. Within 100 Myr, $\sim$ 800 stars with $v_{\mathrm{ej}}$ $\gtrsim$ 50 km s$^{-1}$ are ejected and would be classified as runaway stars. In addition about 30 stellar black holes escape with similar velocities within 100 Myr. Of order 30 stars can be accelerated to high velocities $\sim$ 300 km s$^{-1}$. On average $\sim$ 3000 stars escape at velocities lower than 50 km s$^{-1}$. Overall, the remnants lose $\sim$ 30 percent of their BH population and $\sim$ 3 to 4 per cent of their white dwarf (WD) and star population if MBHs are present. 

In the absence of MBHs the fraction of escaping BHs drops to $\sim$ 6 per cent and to below one per cent for white dwarfs and stars. Comparison simulations of isolated clusters of the same mass as the merger remnants and initialised with central MBH binaries as well as star cluster mergers without MBHs show that the high velocity ejection process is driven by the MBH binaries and not the cluster merger process itself. For merger simulations without MBHs of order $\sim$ 1000 stars and $\sim$ 10 BHs escape the system at velocities below 50 km s$^{-1}$. Comparisons with a single MBH of 1000 $\mathrm{M}_\odot$ in isolated or merging cluster show a smaller population of runway stars at lower
velocities. Recently, \cite{2024A&A...690A.207P} have highlighted the role of star cluster mergers for the formation of runaway stars. Their simulations feature a more realistic hydrodynamical, hierarchical setup and do not include MBHs. When compared to our idealised star cluster mergers without MBHs we find that one of their 'runaway' star group with a velocity of $\sim$ 39 km s$^{-1}$ is consistent with our simulations. As we assume a runaway velocity limit of 50 km s$^{-1}$ they would fall in our low velocity (LV) escaper regime. The second runway group in \citet{2024A&A...690A.207P} has a velocity of $\sim$ 87 km s$^{-1}$. We do not reach such high velocities by cluster mergers alone, As discussed in \citet{2024A&A...690A.207P},  the more realistic and more dynamical setup in their simulation might allow for higher escape velocities during cluster mergers. 

Based on the evolution of the structural and dynamical cluster properties and the MBH binary hardening rates we expect the binary MBHs in our simulations to
merge in less than a Hubble time and produce observable gravitational wave (GW) emission events, detectable by future gravitational wave detectors like Advanced LIGO, the Einstein Telescope, or the Laser Interferometer Space Antenna (LISA). The gravitational recoil kicks exerted on the MBH merger remnants might easily eject the MBHs from the star clusters leaving no detectable evidence for their prior existence apart from a strongly reduced stellar BH sub-population. 

Low mass MBHs might form naturally by stellar collisions in dense stars clusters \citep[see e.g.][and references therein]{2004Natur.428..724P,2021MNRAS.501.5257R,Rantala2024}, in particular in the dense, star forming, low metallicity interstellar medium in the early Universe. 
Mergers of these clusters are very likely during their hierarchical formation, during nuclear star cluster assembly or in the disks of early compact galaxies. In this case, the results of our study  imply that interactions with low mass MBH binaries formed in merging stars clusters are an important additional channel for producing runaway and high-velocity stars as well as free floating stellar BHs and compact objects which are produced at high redshift and still populate the disks and halos of present day galaxies.

\section*{Acknowledgements}

TN acknowledges the support of the Deutsche Forschungsgemeinschaft (DFG, German Research Foundation) under Germany’s Excellence Strategy - EXC-2094 - 390783311 of the DFG Cluster of Excellence ''ORIGINS''.

\section*{SOFTWARE}

\bifrost{} \citep{Rantala2023}, NumPy \citep{Harris_2020}, SciPy \citep{Virtanen_2020}, Matplotlib \citep{Hunter_2007}, pygad \citep{Roettgers_2020}, Seaborn \citep{Waskom_2021}, Pandas \citep{Pandas_2023}.

\section*{Data Availability}

The data underlying this article will be shared on reasonable request
to the corresponding author.


\bibliographystyle{mnras}
\bibliography{example} 


\appendix
\section{Dynamically Formed Binaries in Star \& Globular Clusters}
\label{apndx:A}

The increased densities during the core collapse phase of star clusters elevates the rate of dynamical encounters, leading to the formation of binaries through three-body interactions. The encounter rate $\bar{C}$ of three initially unbound bodies leading to binary formation in a system of equal mass objects $m_*$ is \citep{Goodman_1993,Binney2008}

\begin{equation}
\bar{C} \approx \frac{G^5 m_{\star}^5 n^3}{\sigma^9},
\end{equation}

where $n$ is the number density and $\sigma$ the velocity dispersion of the system. Since the above rate refers to equal-mass bodies only, it is however still uncertain how well it predicts the binary formation rates in multi-component environments such as star and globular clusters and/or galactic nuclei. \cite{Atallah_2024} found that in such encounters, it is highly unlikely that the two most massive bodies become bound, in contrast to what has mostly been  suggested in the literature so far. The large number of single bodies in the system leads to frequent fly-by and exchange interactions with the formed binaries, where one of the member is replaced by a field star. The higher the mass of the incoming body, the higher the probability of it replacing a binary member \citep{Valtonen_2006}. This inevitably results to massive stars becoming members of binaries. 

For a system containing stellar-mass BHs the above mechanisms allow the efficient formation of binary BHs \citep{Portegies_Zwart_2000a,Park_2017,Kremer_2019,Torniamenti_2022,Kritos_2024,Rantala2024}, potentially leading to their GW-driven coalescence, covering a broad mass-ratio spectrum, detectable with current \citep{Abbott_2016,Abbott_2019} and future \citep{Amaro_Seoane_2017_LISA} gravitational wave detectors. Additionally, BH-star binaries can be formed through exchange in 3- or 4-body encounters \citep[e.g.][]{Ryu_2023b}{}{} in star massive star clusters \citep[e.g.][]{Rastello_2023,Pina_2024,Fantoccoli_2024}{}{} potentially explaining the three dormant BHs detected by Gaia \citep{Gaia_2016}: Gaia BH1 \citep{El_Badry_2022}, BH2 \citep{El_Badry_2023} and BH3 \citep{Gaia_2024}.


\bsp	
\label{lastpage}
\end{document}